\documentclass[preprint,aps,12pt,showpacs,nofootinbib,tightenlines]{revtex4}
\usepackage{amsmath}
\usepackage{amssymb}
\usepackage{epsfig}
\usepackage{graphicx}
\begin{document}
\preprint{NJNU-TH-2003-17}
\newcommand{\beq}{\begin{eqnarray}}
\newcommand{\eeq}{\end{eqnarray}}

\newcommand{\bxsga}{B\to X_s \gamma}
\newcommand{\brbxsga}{{\cal B}(B\to X_s \gamma)}
\newcommand{\bzbzb}{ B_d^0 - \bar{B}_d^0 }

\newcommand{\bsga}{  b\to s \gamma}
\newcommand{\bdga}{  b\to d \gamma}
\newcommand{\bvga}{  B\to V \gamma }
\newcommand{\bksga}{ B\to K^* \gamma}
\newcommand{\brhoga}{B\to \rho \gamma}

\newcommand{\brbkz}{{\cal B}(B\to \overline{K}^{*0} \gamma)}
\newcommand{\brbkm}{{\cal B}(B\to K^{*-} \gamma)}
\newcommand{\brbrm}{{\cal B}(B\to \rho^- \gamma)}
\newcommand{\brbrz}{{\cal B}(B\to \rho^0 \gamma)}

\newcommand{\calb}{ {\cal B}}
\newcommand{\acp}{ {\cal A}_{CP}}
\newcommand{\oas}{ {\cal O} (\alpha_s)}

\newcommand{\mt}{m_t}
\newcommand{\mw}{M_W}
\newcommand{\mhp}{M_{H}}
\newcommand{\muw}{\mu_W}
\newcommand{\mub}{\mu_b}
\newcommand{\dmd}{\Delta M_{B_d} }
\newcommand{\ltt}{\lambda_{tt} }
\newcommand{\lbb}{\lambda_{bb} }
\newcommand{\rhob}{\bar{\rho} }
\newcommand{\etab}{\bar{\eta} }

\newcommand{\smallsm}{{\scriptscriptstyle SM}}
\newcommand{\smallyy}{{\scriptscriptstyle YY}}
\newcommand{\smallxy}{{\scriptscriptstyle XY}}
\newcommand{\smallnp}{{\scriptscriptstyle NP}}

\newcommand{\tab}[1]{Table \ref{#1}}
\newcommand{\fig}[1]{Fig.\ref{#1}}
\newcommand{\real}{{\rm Re}\,}
\newcommand{\im}{{\rm Im}\,}
\newcommand{\non}{\nonumber\\ }

\def \epjc{  Eur. Phys. J. C }
\def \jpg{  J. Phys. G }
\def \npb{  Nucl. Phys. B }
\def \plb{  Phys. Lett. B }
\def \prd{  Phys. Rev. D }
\def \prl{  Phys. Rev. Lett.  }
\def \pr{   Phys. Rep. }
\def \rmp{  Rev. Mod. Phys. }
\title{Exclusive $B \to (K^*, \rho)  \gamma$ decays in the general
two-Higgs-doublet models }
\author{ Zhenjun Xiao}
\email{xiaozhenjun@pine.njnu.edu.cn}
\affiliation{Department of Physics, Nanjing Normal University, Nanjing,
Jiangsu 210097, P.R.China}
\affiliation{CCAST(World Laboratory), P.O.Box 8730, Beijing 100080, China}
\author{Ci Zhuang }
\affiliation{Department of Physics, Nanjing Normal University,
Nanjing, Jiangsu 210097, P.R.China}
\date{\today}
\begin{abstract}
By employing the QCD factorization approach, we calculated  the next-to-leading order
new physics contributions to the branching ratios, CP asymmetries, isospin and U-spin
symmetry breaking of the exclusive decays $B \to V \gamma$ ($V=K^*, \rho$),
induced by the charged Higgs penguins in the
general two-Higgs-doublet models. Within the considered parameter space, we
found that
(a) the new physics corrections to  the observables
are generally small in the model I and model III-A, moderate in model II, but large
in model III-B;
(b) from the well measured branching ratios and upper limits,
a lower bound of $\mhp > 200$ GeV in model II was obtained, while the allowed
range of $\mhp$ in model III-B is $ 226 \leq \mhp \leq 293 $ GeV;
these bounds are comparable with those from the inclusive $B \to X_s \gamma$ decay;
(c)the NLO Wilson coefficient $C_7(m_b)$ in model III-B is positive and
disfavered by the measured value of isospin symmetry breaking
$\Delta_{0-}^{exp} (K^*\gamma) = (3.9 \pm 4.8)\%$, but still can not be excluded
if we take the large errors into account;
(d) the CP asymmetry $\acp(B \to \rho \gamma)$ in model III-B has an opposite sign
with the one in the standard model (SM), which may be used as a good observable
to distinguish the SM from model III-B;
(e) the isospin symmetry breaking $\Delta(\rho\gamma)$ is less than $10\%$ in the
region of $\gamma = [ 40 \sim 70]^\circ$ preferred by the global fit result,
but can be as large as $20$ to $40\%$ in the regions of $\gamma \leq 10^\circ$ and
$\gamma \geq 120^\circ$. The SM and model III-B predictions for $\Delta(\rho \gamma)$
are opposite in sign for small or large values of the CKM angles;
(f) the U-spin symmetry breaking $\Delta U(K^*,\rho)$ in the SM and the general
two-Higgs-doublet models is generally small in size: $\sim 10^{-7}$.
\end{abstract}


\maketitle

\newpage

\section{introduction}

As is well known, the inclusive radiative decays $B \to X_q \gamma$ with $q=(d,s)$ and
the corresponding exclusive decays $B \to V \gamma$ ($V=K^*, \rho$) are very sensitive to
the flavor structure of the standard model (SM) and to the new physics models beyond
the SM and have been studied in great detail by many authors\cite{buras96,hurth02,bosch03}.

For the inclusive $B \to X_s \gamma$ decay mode, the world average of the branching ratio
\cite{jessop02}
is
\beq
{\cal B}(B \to X_s \gamma) = (3.34 \pm 0.38) \times 10^{-4}
\eeq
which agrees perfectly with the SM theoretical prediction at the next-to-leading
order (NLO) \cite{cmm97,kagan99,buras02,gm01,greub03}
and puts perhaps most stringent bounds on many new physics models
\cite{carena01,bor00,2hdm,xiao03} where new particles such as the charged Higgs
bosons may provide significant contributions through flavor changing loops.

The exclusive decay $B \to K^* \gamma$ has very clean experimental signal and low
background, which was first observed by CLEO in 1992
\cite{cleo-vg}, and measured recently by BaBar and Belle with good precision
\cite{babar-vg,belle-vg}: the world averages of the CP-averaged branching ratios
are \cite{nakao03}
\beq
{\cal B}(B \to \overline{K}^{*0} \gamma) = (4.17 \pm 0.23) \times 10^{-5}, \non
{\cal B}(B \to K^{*-} \gamma) = (4.18 \pm 0.32) \times 10^{-5},
\label{eq:bvga-exp}
\eeq
and have reached a statistical accuracy of better than $10\%$.
The measurements of the Cabibbo suppressed $B \to (\rho, \omega)\gamma$ decays are difficult
because the signal is about $20$ times smaller, and the continuum background is about
$3$ times larger than the $B \to K^* \gamma$ decay mode. Consequently, experiments have so
far provided only upper bounds\cite{cleo-vg,babar-vg,belle-vg}, but they will be surely
measured at B factories in the near future.  The currently available data as
presented at the LP'2003 conference \cite{nakao03} are summarized in \tab{bvga}.

When compared with the inclusive $B \to X_{s,d} \gamma$ decays, the corresponding exclusive
$B \to V \gamma$ decays are experimentally more tractable (specifically for $B \to \rho \gamma$
mode) but theoretically less clean, since the bound state
effects are essential and need to be described by some no-perturbative quantities
like form factors and light-cone distribution amplitudes (LCDAs).

In Refs.\cite{deshpande87,greub95,aaw99}, the branching ratios and rate asymmetries of
$B \to V \gamma$ decays were investigated in leading order (LO) and next-to-leading
order (NLO) by employing the constituent quark model (CQM) \cite{deshpande87}.
In Ref.\cite{li99}, the exclusive $B \to K^* \gamma$ decay was studied by using
the perturbative QCD approach.
Very recently, in the heavy quark limit $m_b \gg \Lambda_{QCD}$, the decay amplitudes for
the exclusive $B \to (K^*, \rho) \gamma$  decay modes have been calculated in a
model-independent way  by using a QCD factorization approach \cite{beneke01,ali02a,bosch02a}
, which is similar in spirit to the scheme developed earlier for the non-leptonic
two-body decays of B meson \cite{bbns99}.
The NLO standard model predictions for the branching ratios, CP and isospin
asymmetries,as well as the U-spin breaking effects for $B \to K^* \gamma$ and
$B \to \rho \gamma$ decays are now available
\cite{beneke01,ali02a,bosch02a,bosch02b,kagan02}.
The new physics effects on isospin symmetry breaking and direct CP violation
in $B \to \rho \gamma$ decay have also been studied recently in supersymmetric
models \cite{ali00}.

In a previous paper, we calculated the NLO new physics contributions
to the $B^0 - \overline{B}^0$ mixing and the inclusive $\bxsga$ decay
from the charged Higgs loop diagrams in the third type of two-Higgs-doublet model
(model III) and the conventional model II.
In this paper,we calculate the new physics contributions to the branching ratios, CP
asymmetries,  and isospin and U-spin symmetry breaking of the exclusive radiative
decays $B \to (K^*, \rho ) \gamma$ in the
framework of the general two-Higgs-doublet models, including the conventional
models I and II, and the model III. The QCD factorization method for exclusive
$B \to V \gamma$ decays as presented in Refs.\cite{beneke01,ali02a,bosch02a}
will be employed in our calculations.

This paper is organized as follows. In Sec.~II, we describe the basic
structures  of the general two-Higgs-doublet models, give a brief review
about the calculation of $B \to V \gamma$ at NLO in QCD factorization approach in the SM
and present the needed analytical formulas for the calculation of Wilson coefficients and
physical observables.
In Sec.~III and IV, we calculate the NLO new physics contributions to the
$B \to K^* \gamma$ and $B \to \rho \gamma$ decay, respectively.
The conclusions are included in the final section.

\begin{table}[t]
\begin{center}
\caption{Experimental measurements of the CP-averaged branching ratios and/or  CP violating
asymmetries $A_{CP}$ (at $90\% C.L.$) of the exclusive  $B \to V \gamma$ decays for
$V=K^*, \rho$ and $\omega$.}
\label{bvga}
\vspace{0.2cm}
\begin{tabular} {l|l|l|l|l}  \hline
Channel & CLEO\cite{cleo-vg}  & BaBar\cite{babar-vg} & Belle\cite{belle-vg} & Average\\ \hline
$\calb (B \to K^{*0} \gamma) (10^{-5}) $ &$4.55\pm 0.70 \pm 0.34 $ &$4.23 \pm 0.40 \pm 0.22$
                               &$4.09 \pm 0.21 \pm 0.19$ & $4.17 \pm 0.23 $ \\
                             \hline
$\calb(B \to K^{*+} \gamma)(10^{-5}) $ &$3.76\pm 0.86 \pm 0.28$& $3.83 \pm 0.62 \pm 0.22$
                              &$4.40 \pm 0.33 \pm 0.24 $ & $4.18 \pm 0.32 $\\
                            \hline
$\calb(B \to \rho^0 \gamma) (10^{-6})$ & $ < 17 $& $ < 1.2 $
                      & $ < 2.6 $ & \\ \hline
$\calb (B \to \rho^+ \gamma) (10^{-6})$ & $< 13 $ & $< 2.1 $
                      & $< 2.7 $ & \\ \hline
$\calb (B \to \omega \gamma) (10^{-6})$   &  & $< 1.0 $&$ < 4.4 $ &
\\ \hline
${\cal A}_{CP} (B \to K^{*0} \gamma) (\%) $ &$8 \pm 13 \pm 3$
                              &$-3.5 \pm 9.4\pm 2.2$
                              &$-6.1\pm 5.9 \pm 1.8$& \\
$\acp(B \to K^{*+} \gamma) (\%)$  && &$+5.3 \pm 8.3 \pm 1.6 $& \\ \hline
\end{tabular}\end{center}
\end{table}

\section{Theoretical framework  }\label{sec:th}

For the standard model part, we follow the procedure of Ref.\cite{bosch02a} and use the
formulas as presented in Refs.\cite{bosch02a,bosch02b}.
The QCD factorization approach to the exclusive $B \to V \gamma$ decays was applied
independently in Refs.\cite{beneke01,ali02a,bosch02a} with some differences in the
definition and explicit expressions of functions. We adopt the analytical
formulas in the SM as presented in Refs.\cite{bosch02a,bosch02b} in this paper, since more
details can be found there.

In this section, we present the effective Hamiltonian and the relevant formulas
for the exclusive decays $B \to V \gamma $ in the framework of the SM and the
general two-Higgs-doublet models.

\subsection{Effective Hamiltonian for inclusive $b \to s \gamma$ decay}

In the framework of the SM, if we only take into account operators up to dimension 6 and put
$m_s=0$, the effective Hamiltonian for $b \to s \gamma $ transitions  at the
scale $\mu$ reads \cite{bosch02a}
\beq
{\cal H}_{eff} = \frac{G_F}{\sqrt{2}}
\sum_{p=u,c}{\lambda_p^s \left [C_{1}Q_{1}^p+C_{2}Q_{2}^p+\sum_{j=3}^8 C_{j}Q_{j} \right ]}
\label{eq:heff}
\eeq
where $\lambda_p^q=V_{pq}^{*}V_{pb}$ for $q=(d,s)$ is the Cabibbo-Kobayashi-Maskawa (CKM)
factor \cite{ckm}. And the current-current,
QCD penguin,  electromagnetic and chromomagnetic dipole operators in the standard basis
\footnote{There is another
basis: the CMM basis, introduced by Chetyrkin, Mosiak, and M\"unz \cite{cmm97}
where the fully anticommuting $\gamma_5$ in dimensional regularization are employed.
The corresponding operators and Wilson coefficients in the CMM basis are denoted as
$P_i$ and $Z_i$ in \cite{bosch02a}.
More details see \cite{cmm97,bosch02b}. } are given by
\footnote{For the numbering of operators $Q_{1,2}^p$, we use the same convention as
Ref.\cite{bosch02b} throughout this paper.}
\beq
Q_1^p&=&(\bar{s}p)_{V-A}(\bar{p} b)_{V-A}\, , \non
Q_2^p&=&(\bar{s_\alpha}p_\beta)_{V-A}(\bar{p_\beta} b_\alpha)_{V-A}\, ,\non
Q_3&=&(\bar{s}b)_{V-A}\sum(\bar{q}q)_{V-A}\, ,\non
Q_4&=&(\bar{s_\alpha}b_\beta)_{V-A}\sum(\bar{q_\beta} q_\alpha)_{V-A}\, , \non
Q_5&=&(\bar{s}b)_{V-A}\sum(\bar{q}q)_{V+A}\, ,\non
Q_6&=&(\bar{s_\alpha}b_\beta)_{V-A}\sum(\bar{q_\beta}q_\alpha)_{V+A}\, ,\non
Q_7&=&\frac{e}{8\pi^2}m_b\bar{s_\alpha}\sigma^{\mu\nu}(1+\gamma_5)b_\alpha F_{\mu\nu} \, ,\non
Q_8&=&\frac{g}{8\pi^2}m_b\bar{s_\alpha}\sigma^{\mu\nu}(1+\gamma_5)T_{\alpha \beta}^ab_\beta
G_{\mu\nu}^a\,  , \label{eq:qi}
\eeq
where $T_a$ ($a=1,\ldots , 8$) stands for $SU(3)_c$ generators, $\alpha$ and $\beta$
are color indices,
$e$ and $g_s$ are the electromagnetic and strong coupling constants,
 $Q_1$ and $Q_2$ are current-current operators, $Q_3-Q_6$ are the QCD penguin
operators, $Q_7$ and $Q_8$ are the electromagnetic and chromomagnetic penguin operators.
The effective Hamiltonian for $b \to d \gamma $ is obtained from
Eqs.(\ref{eq:heff}) - (\ref{eq:qi}) by the replacement $s \to d$.

To calculate the exclusive $B \to V\gamma$ decays complete to next-to-leading order in
QCD and to leading order in $\Lambda_{QCD}/M_B$, only the NLO Wilson coefficient $C_7(\mub)$
and LO Wilson coefficients
$C_i(\mub)$ with $i=1-6,8$ and $\mub = {\cal O}(m_b)$ are needed.
For the sake of the readers, we simply present these Wilson coefficients at the high
matching scale $\mw$ and the low energy scale $\mub =m_b$ here, one can see
\cite{buras96,cmm97} for more details.

In literature, one usually uses certain linear combinations of
the original $C_i(\mu)$, the so-called ``effective coefficients" $C^{{\rm eff}}(\mu)$
introduced in Refs.\cite{buras94,cmm97}, in ones calculation. The corresponding
transformations are of the form
\beq
C^{{\rm eff}}_i(\mu) &=& C_i(\mu), \ \ \ (i=1,\ldots , 6), \\
C^{{\rm eff}}_7(\mu) &=& C_7(\mu)   + \sum_{i=1}^6 y_i C_i(\mu),  \\
C^{{\rm eff}}_8(\mu) &=& C_8(\mu)   + \sum_{i=1}^6 z_i C_i(\mu),
\eeq
with $\vec{y} = (0,0,0,0,-1/3,-1)$ and $\vec{z}=(0,0,0,0,1,0)$ in the NDR
scheme \cite{buras94}, and $\vec{y} = (0,0,-1/3,-4/9,-20/3,-80/9)$ and
$\vec{z}=(0,0,1,1/6,20,-10/3)$ in the $\overline{MS}$ scheme with fully anticommuting
$\gamma_5$ \cite{cmm97}. In order to simplify the notation we will also omit
the label ``eff" throughout this paper.

Within the SM and at the matching scale $\mu=\mw$, the leading order Wilson coefficients are
\beq
C_{1,\smallsm}^{0}(M_W) &=& 1,\\
C_{i,\smallsm}^{0}(M_W) &=& 0, i=2,...,6\\
C_{7,\smallsm}^{0}(M_W) &=& - \frac{A(x_t)}{2}, \label{eq:c70mw}\\
C_{8,\smallsm}^{0}(M_W) &=& - \frac{D(x_t)}{2}, \label{eq:c80mw}
\eeq
with
\beq
A(x) &=& \frac{3x_{t}^3-2x_{t}^2}{4(x_{t}-1)^4}\ln{x_t}
                     +\frac{-8x_{t}^3-5x_{t}^2+7x_t}{24(x_t-1)^3},\label{eq:ax}\\
D(x) &=& \frac{-3x_{t}^2}{4(x_{t}-1)^4}\ln{x_t}
                     +\frac{-x_{t}^3+5x_{t}^2+2x_t}{8(x_t-1)^3},  \label{eq:dx}
\eeq
while the NLO results for $C_7(\mw)$ and $C_8(\mw)$ are
\beq
 C_{7,\smallsm}^{1}(M_W) &=& \frac{-16x_t^4-122x_t^3+80x_t^2-8x_t}{9(x_t-1)^4}Li_2(1-\frac{1}{x_t})
                +\frac{6x_t^4+46x_t^3-28x_t^2}{3(x_t-1)^5}\ln{x_t}^2\non
                  &&+\frac{-102x_t^5-588x_t^4-2262x_t^3
                  +3244x_t^2-1364x_t+208}{81(x_t-1)^5}\ln{x_t}\non
                 && +\frac{1646x_t^4+12205x_t^3-10740x_t^2+2509x_t-436}{486(x_t-1)^4},
                 \label{eq:c71mw}\\
 C_{8,\smallsm}^{1}(M_W) &=& \frac{-4x_t^4+40x_t^3+41x_t^2+x_t}{6(x_t-1)^4}Li_2(1-\frac{1}{x_t})
                   +\frac{-17x_t^3-31x_t^2}{2(x_t-1)^5}\ln{x_t}^2\non
                  && +\frac{-210x_t^5+1086x_t^4+4893x_t^3+2857x_t^2-1994x_t+208}{216(x_t-1)^5}\ln{x_t}\non
                  &&+\frac{737x_t^4-14102x_t^3-28209x_t^2+610x_t-508}{1296(x_t-1)^4}
\label{eq:c81mw}
\eeq
where $x_t=m_t^2/m_w^2$, and $Li_2(x)$ is the dilogarithm function.

At the low energy scale $\mu =O(m_b)$, the leading order Wilson coefficients  are
\beq
 C_{j,\smallsm}^{0}(\mu) &=& \sum_{i=1}^{8}k_{ji}\eta^{a_i},  \ \  for \ \ j = 1,...,6,\\
 C_{7,\smallsm}^{0}(\mu) &=&
 \eta^{\frac{16}{23}}C_{7,\smallsm}^{0}(M_{W})
 +\frac{8}{3}(\eta^{\frac{14}{23}}-\eta^{\frac{16}{23}})C_{8,\smallsm}^{0}(M_W)
 +\sum_{i=1}^{8}h_i\eta^{a_i}, \label{eq:c70mb}\\
 C_{8,\smallsm}^{0}(\mu) &=& C_{8,\smallsm}^{0}(M_W)\eta^{\frac{14}{23}}
 + \sum_{i=1}^{8}\hbar_8\eta^{a_i}
\label{eq:c80mb}
 \eeq
in the standard basis, while
\beq
 Z_{j,\smallsm}^{0}(\mu) &=& \sum_{i=1}^{8}{h_{ji}\eta^{a_i}}, \ \  for\ \ j = 1...6,\\
 Z_{7,\smallsm}^{0}(\mu) &=& C_{7,\smallsm}^{0}(\mu),\label{eq:z70mb}\\
 Z_{8,\smallsm}^{0}(\mu) &=& C_{8,\smallsm}^{0}(\mu)\label{eq;z80mb}
 \eeq
in the CMM basis.

The NLO Wilson coefficient $C_7(\mub)$ at scale $\mub={\cal O}(m_b)$ can be written as
\beq
C_{7,\smallsm}(\mu) = C_{7,\smallsm}^0(\mu)
+ \frac{\alpha_s(\mu)}{4\pi} C_{7,\smallsm}^1(\mu)\label{eq:c7nlo}
 \eeq
with
\beq
C_{7,\smallsm}^{1}(\mu) &=& \eta^{\frac{39}{23}}C_{7,\smallsm}^{1}(M_{W})+\frac{8}{3}
                 \left (\eta^{\frac{37}{23}}-\eta^{\frac{39}{23}} \right )C_{8,\smallsm}^{1}(M_W)\non
            &&  +\left ( \frac{297664}{14283}\eta^{\frac{16}{23}}
                -\frac{7164416}{357075}\eta^{\frac{14}{23}}
                +\frac{256868}{14283}\eta^{\frac{37}{23}}
                -\frac{6698884}{357075}\eta^{\frac{39}{23}}\right ) C_{8,\smallsm}^{0}(M_W)\non
            &&  +\frac{37208}{4761}\left ( \eta^{\frac{39}{23}}-\eta^{\frac{16}{23}} \right )
            C_{7,\smallsm}^{0}(M_W)
                +\sum_{i=1}^{8}{(e_i\eta E(x_t)+ f_i + g_i\eta)\eta^{a_i}},
\eeq
with
\beq
 E(x) & = & \frac{x (x^2+11x-18)}{12 (x-1)^3}
+\frac{x^2 (4 x^2-16x+15)}{6(x-1)^4} \ln x-\frac{2}{3} \ln x -\frac{2}{3},
\eeq
where $\eta=\alpha_s(\mw)/\alpha_s(\mub)$, and the ``magic numbers" $a_i$,  $e_i$,
$f_i$, $g_i$, $K_{ji}$ and $h_{ji}$, $h_i$ and $\hbar_i$ are summarized in \tab{magic}.

\begin{table}
\begin{center}
\caption{The "magic numbers" appeared in the calculations of the the Wilson coefficients
$C_i(\mu)$ in the rare decay $b \to q \gamma$ with $q=(d,s)$.}
\label{magic}
\begin{tabular}{c c c c c c c c c  } \hline
$i$&1&2&3&4&5&6&7&8\\
\hline\hline
$a_i$&$\frac{14}{23}$&$\frac{16}{23}$&$\frac{6}{23}$&$-\frac{12}{23}$
&0.4086&-0.4230&-0.8994&0.1456\\ \hline
$k_{1i}$&0&0&$\frac{1}{2}$&$ \frac{1}{2}$&0&0&0&0\\
$k_{2i}$&0&0&$\frac{1}{2}$&$-\frac{1}{2}$&0&0&0&0\\
$k_{3i}$&0&0&$-\frac{1}{14}$&$\frac{1}{6}$&0.0510&-0.1403&-0.0113&0.0054\\
$k_{4i}$&0&0&$-\frac{1}{14}$&$-\frac{1}{6}$&0.0984&0.1214&0.0156&0.0026\\
$k_{5i}$&0&0&0&0&-0.0397&0.0117&-0.0025&0.0304\\
$k_{6i}$&0&0&0&0&0.0335&0.0239&-0.0462&-0.0112\\
\hline
$h_{1i}$&0&0&1&-1&0&0&0&0\\
$h_{2i}$&0&0&$\frac{2}{3}$&$\frac{1}{3}$&0&0&0&0\\
$h_{3i}$&0&0&$\frac{2}{63}$&$-\frac{1}{27}$&-0.0659&0.0595&-0.0218&0.0335\\
$h_{4i}$&0&0&$\frac{1}{21}$&$\frac{1}{9}$&0.0237&-0.0173&-0.01336&-0.0136\\
$h_{5i}$&0&0&$-\frac{1}{126}$&$\frac{1}{108}$&0.0094&-0.01&0.001&-0.0017\\
$h_{6i}$&0&0&$-\frac{1}{84}$&$-\frac{1}{36}$&0.0108&0.0163&0.0103&0.0023\\
\hline
$e_i$&$\frac{4661194}{816831}$&$-\frac{8516}{2217}$&0&0&-1.9043&-0.1008&0.01216&0.0183\\
$f_i$&-17.3023&8.5027&4.5508&0.7519&2.0040&0.7476&-0.5358&0.0914\\
$g_i$&14.8088&-10.809&-0.8740&0.4218&-2.9347&0.3971&0.1600&0.0225\\
\hline
$h_{i}$&2.2996&-1.0880&$-\frac{3}{7}$&$-\frac{1}{14}$&-0.6494&-0.0380&-0.0185&-0.0057\\
$\hbar_i$&0.8623&0&$0$&$0$&-0.9135&0.0873&-0.0571&0.0209\\
\hline
\end{tabular}
\end{center}
\end{table}

\begin{table}
\begin{center}
\caption{Values of the input parameters used in the numerical calculations
\cite{pdg2003,ball98,b03}. For the value of $F_{K^*}$, we use the lattice
QCD determination of $F_{K*} =0.25 \pm 0.06$\cite{b03} instead of the result
$F_{K^*}=0.38 \pm 0.06$ as given in Ref.\cite{ball98}. The smaller value of $F_{K^*}$ gives
a better agreement between the SM predictions and the data.
$R_b=\sqrt{\bar{\rho}^2+\bar{\eta}^2}$, and $A, \lambda, \bar{\rho}$ and $\bar{\eta}$ are
the ordinary Wolfenstein parameters of the CKM mixing matrix. }
\label{input}
\begin{tabular}{c c c c c c } \\\hline \hline
$ A $    & $\lambda $ & $R_b  $         & $\gamma$           &$G_F$                           & $\alpha_{em}$\\ \hline
$ 0.854$ & $ 0.2196 $ & $0.39\pm 0.08 $ & $(60 \pm 20)^\circ$&$1.1664\times 10^{-5}GeV^{-2} $ & $ 1/137.036$
\\ \hline \hline
$\alpha_s(M_Z)$& $ m_W  $     & $m_t  $& $\Lambda^{(5)}_{\overline{MS}}$   & $m_c(m_b) $      &$ m_u $ \\ \hline
$0.119$        &$ 80.42$ GeV  & $174.3 $ GeV  & $225$ MeV                         & $1.3\pm 0.2$ GeV &  $4.2$
MeV\\ \hline\hline
$ f_B    $& $\lambda_{B} $      & $m_{B_d} $ &$m_b(m_b) $ &$\tau_{B^+}$ &$ \tau_{B^0}$  \\  \hline
$200$ MeV & $(350 \pm 150)$ MeV & $5.279$ GeV&$ 4.2$ GeV  & $ 1.671 ps$  & $1.537 ps $  \\
\hline \hline
$ F_{K^*}$     & $f_{K^*} $ &$f_{K^*}^{\bot}$ & $m_{K^*}$ &$\alpha_1^{K^*}$ &$\alpha_2^{K^*}$  \\
\hline
$0.25 \pm 0.06$& $230$ MeV  &$185$ MeV        & $894$ MeV &$0.2$ & $0.04$ \\
\hline \hline
$ F_{\rho}$     & $f_{\rho} $ &$f_{\rho}^{\bot}$ & $m_\rho$ &$\alpha_1^{\rho}$ &$\alpha_2^{\rho}$  \\
\hline
$0.29 \pm 0.04$& $200$ MeV  &$160$ MeV        & $770$ MeV &$0$ & $0.2$ \\
 \hline \hline
\end{tabular} \end{center}
\end{table}

Using the central values of the input parameters as given in \tab{input}, we
find the the numerical results of the Wilson coefficients  $C_i(m_b)$ and $Z_i(m_b)$
in the SM
\beq
\overrightarrow{C^{0}}(m_b)&=& \left \{ 1.1167, -0.2670,  0.0120, -0.0274, 0.0078, -0.0340, -0.3212, -0.1519
\right \} , \label{eq:ci0} \\
\overrightarrow{Z^{0}}(m_b)&=&\left \{ -0.5339,  1.0280, -0.0055, -0.0727, 0.0005,  0.0012, -0.3212, -0.1519
\right \} , \label{eq:zi0}
\eeq
at the leading order, and
\beq
C_7(m_b) = \underbrace{-0.3212}_{C^0_7(m_b)}  + \underbrace{0.0112}_{\Delta C_7^{NLO}}
= -0.3100
\eeq
at the next-to-leading order, the second term denotes the NLO QCD correction to $C_7^0(m_b)$.

\subsection{$B \to V \gamma$ decay in the QCD factorization approach }

Based on the effective Hamiltonian for the quark level process $b \to s(d) \gamma$,
one can write down the amplitude for $B \to V \gamma$ and calculate
the branching ratios and CP violating asymmetries once a method is derived for computing
the hadronic matrix elements.One typical numerical result obtained by employing the
constituent quark model \cite{deshpande87}
 is
\beq
\calb (B \to K^* \gamma) \approx  5 \times 10^{-5}
\eeq
at both LO and NLO level \cite{aaw99}. Although this theoretical prediction is
in good agreement with the data numerically, but the hadronic models used in
\cite{deshpande87,greub95,aaw99} did not allow a clear separation of short- and long-distance
dynamics and a clean distinction of model-dependent and model-independent
features. By using the QCD factorization approach \cite{beneke01,ali02a,bosch02a},
one can separate systematically perturbatively calculable hard scattering kernels (
$T_i^I$ and $T_i^{II}$ )  from nonperturbative form factors and universal light-cone
distribution amplitudes of $B$, $K^*$ and $\rho$ mesons.
The higher order QCD corrections can therefore be taken into account consistently.

In this paper, we calculate the new physics contributions
to the exclusive decays $B \to K^* \gamma$ and $B \to \rho \gamma$ in
the general two-Higgs-doublet models by employing the QCD factorization approach.
We will always consider the decay widths or branching ratios
averaged over the charge conjugated modes with an obvious exception of the CP
asymmetries.

In QCD factorization approach, the hadronic matrix elements of the operators
$Q_i$ with $i=1,\ldots, 8$ for $B \to V \gamma$ decays  can be written as \cite{bosch02a}
\beq
\langle V\gamma(\epsilon)|Q_i|\bar B\rangle =
\left[ F^{B\to V}(0)\, T^I_{i} +
\int^1_0 d\xi\, dv\, T^{II}_i(\xi,v)\, \Phi_B(\xi)\, \Phi_V(v)\right] \cdot\epsilon
\label{eq:bvg1}
\eeq
where $\epsilon$ is the photon polarization 4-vector, $F^{B\to V}$ is the form factor describing
$B \to V$ decays, $\Phi_B$ and $\Phi_V$ are the universal and nonperturbative light-cone
distribution amplitudes for B and $V$ meson respectively \footnote{For explicit expressions and
more details about $\Phi_B$ and $\Phi_V$, one can see Refs.\cite{bbns99,beneke01} and
references therein. },
$v$ ( $\bar{v} \equiv 1-v$ ) is the momentum fraction of a quark (anti-quark) inside a
light meson: $l_1^+ = vk^+$ and $l_2^+ = \bar{v} k^+$ while $k^\mu =(k^+, k^-,\vec{k}_{\bot})$
is a four vector in the light-cone coordinator,
$\xi$ describes the momentum fraction of the light spectator quark
inside a B meson: $l^+ = \xi p_B^+$ with $\xi = {\cal O}(\Lambda_{QCD}/m_b)$,
and $T^I_{i}$ and $T^{II}_i$ denote the perturbative short-distance
interactions. The QCD factorization formula (\ref{eq:bvg1}) holds up to
corrections of relative order $\Lambda_{QCD}/m_b$.

In the heavy quark limit, the contributions to the exclusive $B \to V \gamma$ decay can
be classified into three classes\footnote{For more details of various contributions and
the corresponding Feynman loops, see for example Ref.\cite{bosch02b} and references therein.}:
\begin{enumerate}
\item The ``Type-I" or ``hard vertex" contributions include (a) the contribution of
the magnetic penguin operator $Q_7$ described by form factor $F^{B \to V}$, which
is the only contribution to the amplitude
of $B \to V \gamma$ at the LO approximation;  and (b) the $\oas$ contribution to the
hard-scattering kernels $T_i^I$ from four-quark operators $Q_{1\ldots 6}$
and the chromomagnetic penguin operator $Q_8$.

\item
The ``Type-II" or ``hard spectator"  contributions include the $\oas$ contribution to the
hard-scattering kernels $T_i^{II}$ from four-quark operators $Q_{1\ldots 6}$
and the chromomagnetic penguin operator $Q_8$.

\item
The ``Weak annihilation" contribution, which is suppressed by one power $\Lambda_{QCD}/m_b$
when compared with the Type-I and II contributions, and the dominant annihilation amplitudes
can be computed within QCD factorization.

\end{enumerate}

Combining all parts together, the decay amplitude to $\oas$ for exclusive $B \to V \gamma$ decay
takes the form of
\beq
A(\overline{B} \to V \gamma) &=& \frac{G_F}{\sqrt{2}} R_{V}
\langle V\gamma|Q_7|\overline{B} \rangle\, ,
\label{eq:avga}
\eeq
with
\beq
R_{V} =  \lambda_u^{(q)}\,\left [  a_7^u(V\gamma) + a_{ann}^u (V\gamma) \right ]
+ \lambda_c^{(q)}\,\left [  a_7^c(V\gamma) +a_{ann}^c (V\gamma) \right ],
\label{eq:rv}
\eeq
where $q=s$ for $V=K^*$, $q=d$ for $V=\rho$, and $a_7^p$ ($p=u,c$) denote the hard vertex
and hard spectator NLO contributions
\beq
a^p_7(V\gamma) &=& C_7(\mu) + \frac{\alpha_s(\mu) C_F}{4\pi}
\left[ \sum_{i=1,2} Z^0_i(\mu) G_i(z_p)+ \sum_{j=3\ldots 6,8} Z^0_j(\mu) G_j\right ] \non
&& \quad\,\,+\frac{\alpha_s(\mu_h) C_F}{4\pi}
\left [ C^0_1(\mu_h) H^V_1(z_p) +\sum_{j=3\ldots 6,8} C^0_j(\mu_h) H^V_j\right ],
\label{eq:a7vga}
\eeq
where $z_q = m_q^2/m_b^2$, $\mu_h=\sqrt{0.5 \mu}$, $C_F=4/3$,
the Wilson coefficients can be found in previous subsection,
the explicit expressions of the functions
$G_i$ and $H^V_j$ can be found in Ref.\cite{bosch02b} and in Appendix \ref{app:gi}.
The functions $a_{ann}^u$ and $a_{ann}^c$ in above equation denote the weak annihilation
contributions and take the form of \cite{bosch02b}
\beq
a_{ann}^u(\bar K^{*0}\gamma) & = & Q_d \left[ a_4 b^{K^*}
+a_6 \left ( d_v^{K^*} +d^{K^*}_{\bar v} \right )\right],\non
a_{ann}^c(\bar K^{*0}\gamma) & = & a_{ann}^u(\bar K^{*0}\gamma),\non
a_{ann}^u(K^{*-}\gamma)      & = & Q_u \left[ a_1 b^{K^*} +a_4 b^{K^*}
+ a_6 \left( -2 d_v^{K^*} +d^{K^*}_{\bar v}\right )\right],\non
a_{ann}^c(K^{*-}\gamma)      & = & Q_u \left[ a_4 b^{K^*}
+ a_6 \left (-2d_v^{K^*} + d^{K^*}_{\bar v} \right )\right], \label{eq:annk}
\eeq
for $\overline{B} \to K^* \gamma$ decays, and
\beq
a_{ann}^u(\rho^0\gamma)      & = & Q_d \left[- a_2  b^\rho + a_4  b^\rho
+ a_6 \left( d^\rho_v +d^\rho_{\bar v}\right )\right],\non
a_{ann}^u(\rho^-\gamma)      & = & Q_u \left[ a_1 b^\rho + a_4  b^\rho
+a_6 \left ( -2 d_v^\rho + d^\rho_{\bar v}\right )\right],\non
a_{ann}^c(\rho^0\gamma)      & = & Q_d \left[ a_4  b^\rho
+ a_6 \left( d^\rho_v +d^\rho_{\bar v} \right )\right],\non
a_{ann}^c(\rho^-\gamma)      & = & Q_u \left[ a_4 b^\rho
+ a_6 \left ( -2 d_v^\rho + d^\rho_{\bar v}\right ) \right], \label{eq:annrho}
\eeq
for $\overline{B} \to \rho \gamma$ decays, where $Q_u=2/3$ and $Q_d=-1/3$ are the electric charge
of up and down quarks, $a_i$ denote the combinations of LO Wilson coefficients
\beq
a_{1,2} &=& C^0_{1,2} + \frac{1}{3} C^0_{2,1}, \non
a_{4} &=& C^0_{4} + \frac{1}{3} C^0_{3}, \non
a_{6} &=& C^0_{6} + \frac{1}{3} C^0_{5}.
\eeq
And finally the functions $b^V$, $d^V_{v}$ and $d^V_{\bar{v}}$ are \cite{bosch02b}
\beq
  b^V &=& \frac{2\pi^2}{F_V}\frac{f_B m_V f_V}{m_B m_b \lambda_B}, \label{eq:bv}\\
  d^V_{\stackrel{(-)}{v}} &=& -\frac{4 \pi^2}{F_V}\frac{f_B f_V^\perp}{m_B m_b}
\left ( 1 \mp \alpha^V_1 +\alpha^V_2 +\ldots \right ). \label{eq:dv}
\eeq
The values of all parameters appeared in above two equations can be found in \tab{input}.

One special feature of the $B \to \rho \gamma$ decay is that the weak annihilation
can proceed  through the
current-current operator with large Wilson coefficient $C_1$. Although the annihilation
contribution is power-suppressed in $1/m_b$, but it is compensated by the large
Wilson coefficient and the occurrence of annihilation at tree level.

From the decay  amplitude in Eq.(\ref{eq:avga}), it is straightforward to write down
the branching ratio for $\overline{B} \to V \gamma$ decay
\beq
\calb (\overline{B}\to V\gamma) = \tau_B\frac{G_F^2\alpha m_B^3m_b^2}{32\pi^4}
\left (1-\frac{m_V^2}{m_B^2} \right )^3
\left | R_{V} \right |^2 c_V^2 |F_V|^2,\label{eq:br-vga}
\eeq
where function $R_V$ has been given in Eq.(\ref{eq:rv}), and $c_V=1$ for $V=K^*, \rho^-$
and $c_V=1/\sqrt{2}$ for $V=\rho^0$. The branching ratios for the CP-conjugated
$B \to V \gamma$ decay are obtained by the replacement of $\lambda_p^{(q)} \to
\lambda_p^{(q)*}$ in function $R_V$.

\subsection{Outline of the general 2HDM's}

The simplest extension of the SM is the so-called two-Higgs-doublet models
\cite{2hdm}. In such models, the tree
level flavor changing neutral currents are absent if one
introduces an {\it ad hoc} discrete symmetry to constrain the 2HDM
scalar potential and Yukawa Lagrangian. Let us consider a Yukawa
Lagrangian of the form\cite{atwood97}
\beq
{\cal L}_Y &=& \eta^U_{ij}\bar{Q}_{i,L} \tilde{\phi_1}U_{j,R} +
\eta^D_{ij}\bar{Q}_{i,L} \phi_1 D_{j,R} +\xi^U_{ij}\bar{Q}_{i,L}
\tilde{\phi_2}U_{j,R} +\xi^D_{ij}\bar{Q}_{i,L} \phi_2 D_{j,R}+
H.c., \label{leff}
\eeq
where $\phi_{i}$ ($i=1,2$) are the two Higgs doublets, $\tilde{\phi}_{1,2}=
i\tau_2 \phi^*_{1,2}$, $Q_{i,L}$ ($U_{j,R}$) with $i=(1,2,3)$ are
the left-handed isodoublet quarks (right-handed   up-type quarks),
$D_{j,R}$  are the right-handed  isosinglet  down-type quarks,
while $\eta^{U,D}_{i,j}$  and $\xi^{U,D}_{i,j}$ ($i,j=1,2,3$ are
family index ) are generally the nondiagonal matrices of the
Yukawa coupling. By imposing the discrete symmetry
\beq
\phi_1 \to - \phi_1, \phi_2 \to \phi_2, D_i \to - D_i, U_i \to  \mp U_i
\eeq
one obtains the so called Model I and Model II.

In model III \cite{hou92,atwood97}, the third type of two-Higgs-doublet models,
no discrete symmetry
is imposed and both up- and down-type quarks then may have
diagonal and/or flavor changing couplings with $\phi_1$ and
$\phi_2$. As described in \cite{atwood97}, one can choose a suitable basis to
express two Higgs doublet $\phi_1$ and $\phi_2$ and define the mass eigenstates
$(H^\pm, \overline{H}^0,h^0, A^0)$. After the rotation of quark
fields, the Yukawa Lagrangian of quarks are of the form
\cite{atwood97},
\beq
{\cal L}_Y^{III} = \eta^U_{ij}\bar{Q}_{i,L}
\tilde{\phi_1}U_{j,R} + \eta^D_{ij}\bar{Q}_{i,L} \phi_1 D_{j,R}
+\hat{\xi}^U_{ij}\bar{Q}_{i,L} \tilde{\phi_2}U_{j,R}
+\hat{\xi}^D_{ij}\bar{Q}_{i,L} \phi_2 D_{j,R} + H.c.,
\label{lag3}
\eeq
where $\eta^{U,D}_{ij}$ correspond to the diagonal mass
matrices of up- and down-type quarks, while the neutral and
charged flavor changing couplings will be \cite{atwood97}
\beq
\hat{\xi}^{U,D}_{neutral}= \xi^{U,D}, \ \ \hat{\xi}^{U}_{charged}=
\xi^{U}V_{CKM}, \ \ \hat{\xi}^{D}_{charged}= V_{CKM} \xi^{D},
\label{cxiud}
\eeq
with
\beq
\xi^{U,D}_{ij}=\frac{ g\,\sqrt{m_im_j}}{\sqrt{2}\mw } \lambda_{ij},
\label{lij}
\eeq
where $V_{CKM}$ is the CKM mixing matrix \cite{ckm}, $i,j=(1,2,3)$
are the generation index. The coupling constants $\lambda_{ij}$ are  free parameters
to be determined by experiments, and they may also be complex.

The two-Higgs doublet models have been studied extensively in literature
at LO and NLO level
\cite{lo2hdm,bg98,ciuchini98,crs98,gm01,hou92,atwood97,aliev99,chao99,wu99,xiao01,xiao03}
and tested experimentally \cite{pdg2003}.
For the model I, the new physics corrections to physical observables  are usually very
small and less interesting phenomenologically.
The model II, however, has been very popular,  since it is the building block of the
minimal supersymmetric standard model  and may provide large contributions to
the mixing and decay processes of $K$ and B meson systems.
The most stringent constraint on model II
may come from the inclusive $B \to X_s \gamma$ decay.  From the experimental measurements and
currently available studies at NLO level \cite{bg98,ciuchini98,crs98,gm01,xiao03},
one get to know the following main features of the conventional models I and  II,
and the model III
\begin{itemize}

\item
For the model I, no bound on $\mhp$ can be obtained from $B \to X_s \gamma$
\cite{gambino01}, since the charged Higgs loops interfere destructively with the SM penguin
diagrams  and decouple for large $\tan{\beta}$.

\item
In model II, the charged Higgs penguins interfere constructively with  their SM counterparts,
and thus always enhance the branching ratio $\brbxsga$. The measured mass splitting
$\dmd = 0.502 \pm 0.007 ps^{-1}$ and the decay rate
$\calb(B \to X_s \gamma) = (3.34 \pm 0.38)\times 10^{-4} $ leads to strong
bounds on both the $\tan{\beta} =v_2/v_1$ and the mass $\mhp$ \cite{gambino01,xiao03}.
The typical bounds at NLO level as given for example in Ref. \cite{xiao03} are
\beq
\tan{\beta} > 0.6
\eeq
and
\beq
\mhp \gtrsim  300 GeV
\eeq
for any value of $\tan{\beta}$, and the $\tan{\beta}$ dependence of the lower bound
saturates for $\tan{\beta} \gtrsim 5$.  This NLO lower bound on $\mhp$ is much stronger than the
direct experimental bound $\mhp > 78.6$ GeV \cite{pdg2003} and the bound from other
observables, such as $R_b$ and $B \to \tau$ decays \cite{gm01}.

\item
For the model III \footnote{In this paper,  the term model III always means the scenario of
a general model III as presented in Ref.\cite{chao99}.
In such model III \cite{chao99}, only the couplings $\lambda_{tt}$ and $\lambda_{bb}$ remain
non-zero, and only the charged Higgs  boson penguin diagram provide a new physics contribution
to $b\to s \gamma$ decay at one loop level. For more details see Refs.\cite{chao99,xiao03}.},
the charged-Higgs loop diagrams can provide significant
contributions to $B^0-\overline{B}^0$ mixing, the inclusive $B \to X_s \gamma$
decay and many other physical observables \cite{atwood97,chao99}. In a previous
paper \cite{xiao03}, we calculated the charged-Higgs contributions to
the  mass splitting $\dmd$ and the decay rate $\brbxsga$ at the NLO level, and found
the  strong constraints on free parameters $\ltt$, $\lbb$ and $\mhp$ from the well measured
$\dmd$ and $\brbxsga$. Two typical choices of $(\ltt, \lbb)$ and the corresponding
constraint on $\mhp$
obtained from the measured branching ratio of $B \to X_s \gamma$  decay  are
\beq
{\rm III-A: } \ \ (\ltt, \lbb)= (0.5, 1), \ \ \mhp > 150 {\rm GeV}
\label{eq:seta}
\eeq
as shown in Fig.9 of Ref.\cite{xiao03}; and
\beq
{\rm III-B: } \ \ (\ltt, \lbb)= (0.5, 22), \ \ 226 \leq \mhp \leq 285 {\rm
GeV}. \label{eq:setb}
\eeq
For the first case, the new physics contribution to $\bxsga$ is very small
and become negligible for $\mhp > 250$ GeV. For the second case (in \cite{xiao03},
it was denoted as the case C), the new
physics contribution can be rather large, the sign of the dominant Wilson
coefficient $C^{eff}_7(m_b)$ changed its sign from negative to positive due to the
inclusion of the charged-Higgs penguin contributions.
In this paper, we denote these two typical cases as the model III-A and III-B,
respectively.

\end{itemize}

\subsection{NLO Wilson coefficients in the  general 2HDM's}

The new physics contributions to the quark level $b \to s/d  \gamma$ transition
from the charged Higgs penguins manifest themselves from the correction to the
Wilson coefficients at the matching scale $\mw$. In Ref.\cite{bg98}, the
authors calculated the NLO QCD corrections  to the $B \to X_s \gamma$ decay in
the conventional models I and II. In Ref.\cite{xiao03}, we extended their work
to the case of model III. Here we firstly present the Wilson coefficients at
the energy scales $\mw$ and $\mu = {\cal O}(m_b)$ in a general 2HDM and then
calculate the branching ratios, CP and isospin asymmetries, and the U-spin symmetry
breaking of the exclusive decays $B \to K^* \gamma$ and $B \to \rho \gamma$
in the following sections.

Note that the CMM basis was used in Refs.\cite{bg98,xiao03}, the Wilson coefficients
$C_i^{eff}(\mu)$ there are indeed the Wilson coefficients $Z_i(\mu)$ in this paper.
For the exclusive decays $B \to V \gamma$ and to the first order in $\alpha_s$,
only the NLO expression for $C_7(\mu)$ has to be used while the leading order values
are sufficient for the other Wilson coefficients appeared in $a_7^p(V\gamma)$
in Eq.(\ref{eq:a7vga}).  Therefore, only $C_7(\mu)$, $C_8^{(0)}(\mu)$ and $Z_8^{(0)}(\mu)$
in Eq.(\ref{eq:a7vga}) are affected by the charged-Higgs penguin contributions, while
all other Wilson coefficients for $i=1,\ldots, 6$ remain the same as in the SM.
Since $Z^{0}_{7,8}(\mu)=C^{0}_{7,8}(\mu)$, and $C_7^{1}(\mu) = Z_7^{1}(\mu)$
\cite{bosch02b}, so we here use the terms $C^{0}_{7,8}$ and $C_7^{1}$ for convenience.

The new physics part of the LO Wilson coefficients $C^{0}_{7,8}$ at the matching energy
scale $\mw$ take the form,
\beq
 C^{0}_{7,\smallnp}(\mw)  & = & -\frac{1}{6} |Y|^2 A(y_t) + (XY^*) B(y_t)\, , \label{eq:c70mw-2}  \\
 C^{0}_{8,\smallnp}(\mw)  & = & -\frac{1}{6} |Y|^2 D(y_t) + (XY^*) E(y_t)\, , \label{eq:c80mw-2}
\eeq
where $y_t=m_t^2/\mhp^2$, and the functions $A(x)$ and $D(x)$ have been given in
Eqs.(\ref{eq:c70mw},\ref{eq:c80mw}), while
\beq
B(y) & = &\frac{3y-5y^2}{12(1-y)^2} + \frac{2y-3y^2}{6(1-y)^3}\log[y],\label{eq:c70-xy}\\
E(y) & = &\frac{3y-y^2}{4(1-y)^2}   + \frac{y}{2(1-y)^3}\log[y].      \label{eq:c80-xy}
\eeq

The new physics parts of the NLO Wilson coefficients $C^{1}_{7,8}$  at the matching
scale $\mu_W$ can be written as
\beq
 C_{7, \smallnp}^{1}(\mw) & = & |Y|^2 \, C_{7,\smallyy}^{1}(\mw)
  +        (XY^*) \, C_{7,\smallxy}^{1}(\mw), \label{eq:c71mw-2}  \\
 C_{8, \smallnp}^{1}(\mw) & = & |Y|^2 \, C_{8,\smallyy}^{1}(\mw)
  +       (XY^*) \, C_{8,\smallxy}^{1}(\mw) \,,
\label{eq:c81mw-2}
\eeq
with
\beq
C_{i,\smallyy}^{1}(\mw) & = &  W_{i,\smallyy} + M_{i,\smallyy} \ln\left [y_t\right ] \,
+ T_{i,\smallyy}  \left( \ln \left [x_t\right ] - \frac{4}{3} \right ) \, , \label{eq:ci-yy}\\
C_{i,\smallxy}^{1}(\mw) & = &  W_{i,\smallxy} + M_{i,\smallxy} \ln[y_t]
                 \,+ T_{i,\smallxy} \left( \ln[x_t] - \frac{4}{3} \right) \,.
\label{eq:ci-xy}
\eeq
The explicit expressions of functions $W_{i,j}$, $M_{i,j}$ and $T_{i,j}$
($i=7,8$ and $j=YY,XY$ ) can be found in Refs.\cite{bg98} or in Appendix B.
The $T_{ij}$ terms appear when expressing $\overline{m}_t(\mw)$ in terms of the pole mass
$m_t$ in the corresponding lowest order coefficients \cite{bg98}.

At low energy scale $\mu = {\cal O}(m_b)$, the Wilson coefficients $C^{0,1}_7(\mu)$ and
$C^0_8(\mu)$ after the inclusion of new physics contributions can be written as
\beq
 C^{0}_7(\mu)  & = &  \eta^\frac{16}{23}
 \left [ C^{0}_{7,\smallsm}(\mw) +C^{0}_{7,\smallnp}(\mw)\right ]\non
&&  +\frac{8}{3} \left (\eta^\frac{14}{23} -\eta^\frac{16}{23}\right )
                  \left [  C^{0}_{8, \smallsm}(\mw) + C^{0}_{8,\smallnp}(\mw)\right ]
 + \sum_{i=1}^8 h_i \,\eta^{a_i}\,, \label{eq:c70mb-2}\\
 C^{0}_8(\mu) & = &   \eta^\frac{14}{23}
 \left [  C^{0}_{8,\smallsm}(\mw) + C^{0}_{8,\smallnp}(\mw)\right ]
 + \sum_{i=1}^8 \hbar_i\,\eta^{a_i}\, , \label{eq:c80mb-2}\\
 C^{1}_7(\mub) & = &  \eta^{\frac{39}{23}}
 \left [  C^{1}_{7, \smallsm}(\mw) + C^{1}_{7,\smallnp}(\mw)\right ]\non
&& + \frac{8}{3} \left( \eta^{\frac{37}{23}} - \eta^{\frac{39}{23}} \right)
     \left [  C^{1}_{8, \smallsm}(\mw) + C^{1}_{8,\smallnp}(\mw)\right ] \non
&& +\left( \frac{297664}{14283}   \eta^{\frac{16}{23}}
       -\frac{7164416}{357075} \eta^{\frac{14}{23}}
       +\frac{256868}{14283}   \eta^{\frac{37}{23}}
       -\frac{6698884}{357075} \eta^{\frac{39}{23}}  \right)\non
&&  \cdot \left [  C^{0}_{8, \smallsm}(\mw)  + C^{0}_{8,\smallnp}(\mw)\right ]\non
&&  +\, \frac{37208}{4761} \left(\eta^{\frac{39}{23}} -\eta^{\frac{16}{23}} \right)
         \left [  C^{0}_{7, \smallsm}(\mw) + C^{0}_{7,\smallnp}(\mw)\right ] \non
&& + \sum_{i=1}^8  \left[ e_i\eta E(x_t)+ f_i + g_i\eta)\eta^{a_i}\right ] \eta^{a_i} \,,
\label{eq:c71mb-2}
\eeq
where the ``magic numbers"  are listed in Table \ref{magic}.

In the conventional model I and II,  the general Yukawa couplings $X$ and $Y$ are real
and given by
\beq
X&=& -\cot{\beta}, \ \ Y=\cot{\beta} \ \ {\rm (Model \ \  I)}\, , \\
X&=& \tan{\beta}, \ \ \ \ Y=\cot{\beta} \ \ \ {\rm (Model \ \ II)}\, .
\label{eq:xy-m2}
\eeq
In the Model III where only the couplings $\ltt$ and $\lbb$ are non-zero,
the relation between the couplings $(X,Y)$ and $(\ltt,\lbb)$ is also simple
\beq
X= -\lbb, \ \ Y=\ltt \ \ {\rm (Model \ \ III)}\, .
\label{eq:xy-m3}
\eeq

Now we are ready to calculate the numerical results for the $B \to V \gamma$ decay
in the general 2HDM's with the inclusion of NLO QCD corrections.

\section{$B \to K^* \gamma$ decay }\label{sec:bks}

For the numerical calculations, unless otherwise specified, we use the central values
of the input parameters as listed in \tab{input},
and consider the uncertainties of those parameters as given explicitly in \tab{input}.

From Eqs.(\ref{eq:avga}) and (\ref{eq:br-vga}), the decay amplitude and branching
ratio for $B \to K^* \gamma$ decay can be written as
\beq
A(\overline{B} \to K^* \gamma) &=& \frac{G_F}{\sqrt{2}} R_{K^*}
\langle K^* \gamma|Q_7|\overline{B} \rangle\, , \label{eq:avks}\\
\calb (\overline{B}\to K^*\gamma) &=& \tau_B\frac{G_F^2\alpha m_B^3m_b^2}{32\pi^4}
\left (1-\frac{m_{K^*}^2}{m_B^2} \right )^3
\left | R_{K^*} \right |^2  |F_K^*|^2,\label{eq:br-ks}
\eeq
with
\beq
R_{K^*} =  V_{us}^* V_{ub}\,\left [  a_7^u(K^*\gamma) + a_{ann}^u (K^*\gamma) \right ]
+ V_{cs}^* V_{cb}\,\left [  a_7^c(K^* \gamma) +a_{ann}^c (K^*\gamma) \right ].
\label{eq:rks}
\eeq

The CP asymmetry of $B \to K^* \gamma$ can also be defined as \cite{babar-vg,belle-vg}
\beq
\acp(K^* \gamma) =\frac{\Gamma (B \to K^* \gamma)
-\Gamma ( \overline{B} \to \overline{K}^* \gamma)}{
\Gamma( B \to K^* \gamma) + \Gamma(\overline{B} \to \overline{K}^* \gamma)}
\label{eq:acp-vks}
\eeq

Another physical observable for $B \to V \gamma$ decay is the isospin symmetry
breaking in the $K^{*\pm} -\overline{K}^{*0}$ or $\rho^\pm - \rho^0$ system.
Since the branching ratios of both $B^- \to K^{*-} \gamma$ and
$\overline{B}^0 \to \overline{K}^{*0} \gamma$ decays have been measured,
the study of the isospin breaking in $B \to V \gamma$ decays becomes very
interesting now \cite{ali00,kagan02}. Following Ref.\cite{kagan02}, the
breaking of isospin symmetry in the $K^{*-} -\overline{K}^{*0}$ system can be defined as
\beq
\Delta_{0-}(K^* \gamma) \equiv \frac{ \eta_{\tau}\brbkz - \brbkm }{
\eta_\tau \brbkz + \brbkm }. \label{eq:iso}
\eeq
where $\eta_\tau =\tau_{B^+}/\tau_{B^0}$, and the CP-averaged branching ratios are understood.

By using the world averages as given in Eq.(\ref{eq:bvga-exp}) and
the ratio $\tau_{B^+}/\tau_{B^0} = 1.083 \pm 0.017 $ \cite{pdg2003}, we
find numerically that
\beq
\Delta_{0-}(K^* \gamma)^{exp} = (3.9 \pm 4.8) \% , \label{eq:d0m-exp}
\eeq
where the errors from the two measured branching ratios and the ratio $\tau_{B^+}/\tau_{B^0} $
have been added in quadrature.
The measured value of isospin symmetry breaking is indeed small as expected previously
\cite{babar-vg,belle-vg}.
Any new physics contribution producing large isospin breaking for $B \to K^* \gamma $
decays will be strongly constrained by this measurement.

\subsection{Branching ratios and CP asymmetries}

By using the formulas  as given in Eqs.(\ref{eq:a7vga}) and (\ref{eq:annk}) and the
central values of input parameters in \tab{input}, we find the SM predictions for
$a_7^{p}(K^*\gamma)$ and $a_{ann}(K^*\gamma)$ at the low energy scale $\mu = m_b$,
\beq
a_7^{u}(K^* \gamma)&=&\underbrace{-0.3212}_{C^0_{7,\smallsm}(m_b)}
 +\underbrace{0.0113}_{\Delta C^1_{7,\smallsm} }
 \; \underbrace{-0.1407-0.0683i}_{T^I-contribution}
 \; \underbrace{0.0330-0.0002i}_{T^{II}-contribution}\non
                    &=&-0.4177-0.0685i, \label{eq:a7ukmb}\\
 a_7^c(K^*\gamma)&=&\underbrace{-0.3212}_{C^0_{7,\smallsm}(m_b) }
 +\underbrace{0.0113}_{\Delta C^1_{7,\smallsm} }
 \; \underbrace{-0.0802-0.0131i}_{T^I-contribution}
 \; \underbrace{-0.0161-0.0120i}_{T^{II}-contribution}\non
                 &=&-0.4063-0.0251i,\label{eq:a7ckmb}\\
a^u_{ann}(\overline{K}^{*0}\gamma) &=& a^c_{ann}(\overline{K}^{*0}\gamma) =-0.0092, \\
a^u_{ann}(          K^{*-}\gamma)  &=& 0.1933, \\
a^c_{ann}(          K^{*-}\gamma)  &=& 0.0046.
\eeq
It is easy to see that (a) the type-I contribution is about 4 times larger than the
type-II contribution, and (b) only the weak annihilation factor $a^u_{ann}(K^{*-} \gamma)$
contributes to the decay $B \to K^* \gamma$ effectively, since for $b \to u $ transition
the power suppression is compensated by the large Wilson coefficient $C_1$ and the
occurrence of annihilation at tree level.

The corresponding NLO SM predictions for branching ratio $\calb (B \to K^*\gamma)$ are
\beq
\calb (B \to \overline{K}^{*0}\gamma )^{SM} &=& \left [ 3.35^{+1.62}_{-1.30}(F_{K^*})
^{+0.57}_{-0.60} (\mu) ^{+0.27}_{-0.10}(\lambda_B) \pm 0.20 (m_c)\right ] \times 10 ^{-5} \non
&=& \left ( 3.35 ^{+1.75}_{-1.45}  \right ) \times 10^{-5}, \label{eq:brk0sm-s}\\
\calb (B \to            K^{*-}\gamma )^{SM} &=&\left [ 3.25 ^{+1.67}_{-1.33}(F_{K^*})
^{+0.25}_{-0.47} (\mu) ^{+0.35}_{-0.14}(\lambda_B) \pm 0.20(m_c)\right ] \times 10 ^{-5} \non
&=& \left ( 3.25 ^{+1.74}_{-1.43}  \right ) \times 10^{-5}, \label{eq:brkpsm-s}
\eeq
where the four major errors have been added in quadrature. The uncertainty
of the form factor $F_{K^*}$ dominate the theoretical error, and the remaining errors
from other input parameters are negligibly small. Although the central values of the
SM predictions for the decay rates are smaller than the world average as given
in Eq.(\ref{eq:bvga-exp}), but they are in good agreement within $1\sigma$
theoretical error. The effect of annihilation contribution on the decay
rates is less than $5\%$ numerically.

If we use $F_{K^*}=0.38 \pm 0.06$ obtained from the light-cone sum rule (LCSR)
\cite{ball98} instead of $F_{K^*}=0.25 \pm 0.06$ in
numerical calculation, we find
\beq
\calb (B \to \overline{K}^{*0}\gamma )^{SM}
&=& \left ( 7.27 ^{+2.58}_{-2.37}  \right ) \times 10^{-5}, \label{eq:brk0sm-h}\\
\calb (B \to            K^{*-}\gamma )^{SM}
&=& \left ( 7.31 ^{+2.57}_{-2.37}  \right ) \times 10^{-5}. \label{eq:brkpsm-h}
\eeq
Here the central values are much larger than the measured values as given in
Eq.(\ref{eq:bvga-exp}),  but still agree with the data within $2\sigma$ errors because of
the large theoretical error. For the purpose to study the new physics
contributions to the exclusive decays $B \to V \gamma$, one prefers a better
agreement between the SM predictions and the high precision data. Therefore, we will
use $F_{K^*}=0.25 \pm 0.06$ in this paper, unless otherwise specified.

For the model I, the theoretical predictions for branching ratios are
\beq
\calb (B \to \overline{K}^{*0}\gamma )^{I} &=&
\left (3.35 ^{+1.75}_{-1.45} \right ) \times 10 ^{-5}, \label{eq:brk0-m1}\\
\calb (B \to            K^{*-}\gamma )^{I} &=&
\left ( 3.25 ^{+1.74}_{-1.43} \right ) \times 10 ^{-5}, \label{eq:brkp-m1}
\eeq
for $\tan{\beta}=4$ and $\mhp =200$ GeV. The $\mhp$
dependence of the branching ratio $\calb (B \to K^* \gamma )$ is very weak:
it will change by less than $2\%$ in the range of $200 \leq \mhp \leq 600 $ GeV.

\begin{figure}[htb] 
\vspace{2cm}
\centerline{\mbox{\epsfxsize=14cm\epsffile{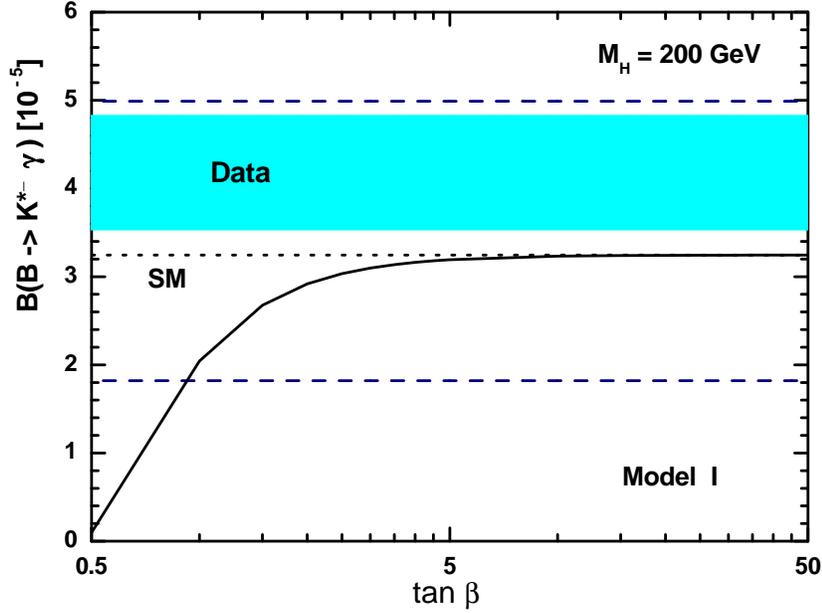}}}
\vspace{-4.5cm}
\caption{ Plot of the branching ratio $\calb (B \to K^{*-} \gamma)$ vs
$\tan{\beta}$ in model I for $\mhp = 200$ GeV.
The dots and solid line shows the central value of  the SM and model I prediction,
respectively. The region between two dashed lines shows the SM prediction:
$\calb (B \to K^{*-} \gamma) = ( 3.25^{+1.74}_{-1.43} )\times 10^{-5}$.
The shaded band shows the data within $2\sigma$ errors:
$\calb (B \to K^{*-} \gamma)^{exp} = ( 4.18 \pm 0.64 )\times 10^{-5}$. }
\label{fig:fig1}
\end{figure}

Fig.\ref{fig:fig1} shows the $\tan{\beta}$ dependence of the branching ratio
$\brbkm$ in model I for $\mhp=200$ GeV.
The dots and solid line shows the central value of  the NLO SM and model I prediction,
respectively. The region between two dashed lines shows the NLO SM prediction with error
as given in Eq.(\ref{eq:brkpsm-s}). The shaded band shows the data:
$\calb (B \to K^{*-} \gamma )^{exp}  = ( 4.18 \pm 0.32 )\times 10^{-5}$.
From this figure, one can see that (a) the NLO SM prediction agree with the
data within $1\sigma$ error; and (b) the new physics contribution in model I
is negligibly small for $\tan{\beta} \geq 1$, while a $\tan{\beta} < 0.5$ is
also strongly disfavored. For $B \to K^{*0} \gamma$ decay mode, we have the
same conclusion.

In the popular model II, the numerical results for $a_7^{p}(K^*\gamma)$
at the low energy scale $\mu = m_b$ are,
\beq
a_7^{u}(K^* \gamma)^{\rm II}&=&\underbrace{-0.3100}_{C_{7,\smallsm}(m_b)}
 -\underbrace{0.06523}_{\Delta C_{7,\smallnp} }
 \, \underbrace{-0.1436-0.0724i}_{T^I-contribution}
 \, \underbrace{+0.0481-0.0003i}_{T^{II}-contribution}\non
                    &=&-0.4707-0.0728 i, \label{eq:a7ukmb-m2}\\
 a_7^c(K^*\gamma)^{\rm II}&=&\underbrace{-0.3100}_{C_{7,\smallsm}(m_b) }
 -\underbrace{0.0652}_{\Delta C_{7,\smallnp} }
 \, \underbrace{-0.0831-0.0172i}_{T^I-contribution}
 \, \underbrace{-0.0265-0.0182i}_{T^{II}-contribution}\non
                 &=&-0.4848-0.0354i,\label{eq:a7ckmb-m2}
\eeq
for $\tan{\beta}=4$ and $\mhp =300$ GeV.
The second terms in above two equations are the new physics corrections to the NLO
Wilson coefficient
$C_{7,\smallsm}(m_b)$, the hard vertex and hard spectator contributions are also changed
slightly because of the variations of $Z_8^{0}(\mu)$ and $C_7^{0}(\mu_h)$
after including the charged-Higgs contributions.  The total new physics contribution
to $a_7^{p}$ in model II is around $10\%$ for  $\tan{\beta}=4$ and $\mhp =300$ GeV.
The annihilation parts remain unchanged.

For the model II, the theoretical predictions for branching ratios are
\beq
\calb (B \to \overline{K}^{*0}\gamma )^{II} &=& \left [ 4.54^{+2.22}_{-1.77}(F_{K^*})
^{+0.68}_{-0.72} (\mu) ^{+0.33}_{-0.13}(\lambda_B) \pm 0.22 (m_c)\right ] \times 10 ^{-5} \non
&=& \left ( 4.54 ^{+2.36}_{-1.93}  \right ) \times 10^{-5}, \label{eq:brk0-m2}\\
\calb (B \to            K^{*-}\gamma )^{II} &=&\left [ 4.47 ^{+2.29}_{-1.83}(F_{K^*})
^{+0.32}_{-0.57})(\mu) ^{+0.44}_{-0.17}(\lambda_B) \pm 0.23(m_c)\right ] \times 10 ^{-5} \non
&=& \left ( 4.47 ^{+2.36}_{-1.94}  \right ) \times 10^{-5}, \label{eq:brkp-m2}
\eeq
for $\tan{\beta}=4$ and $\mhp =300$ GeV.

\begin{figure}[htb]  
\vspace{3cm}
\centerline{\mbox{\epsfxsize=14cm\epsffile{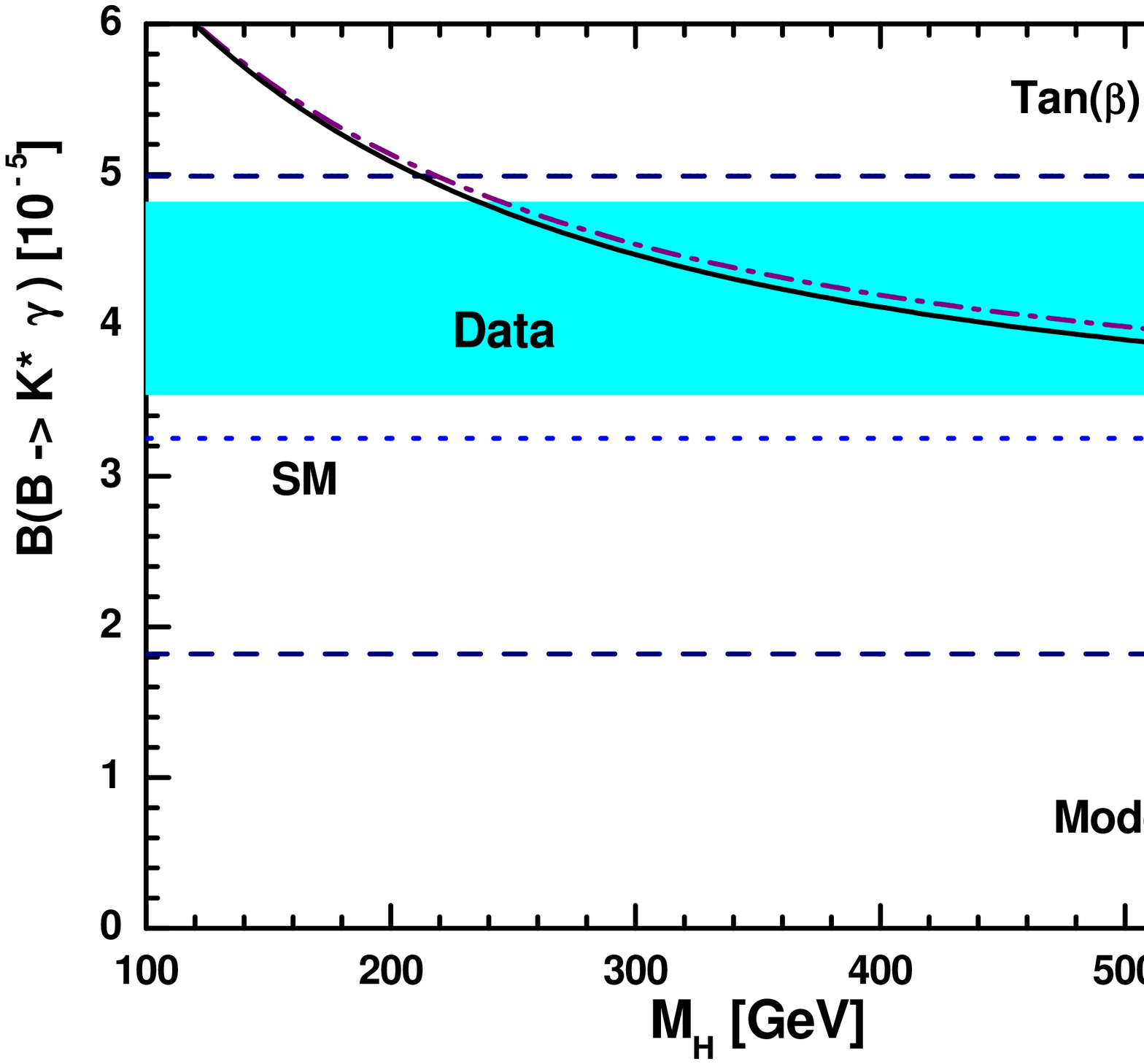}}}
\vspace{-4.5cm}
\caption{ The $\mhp$ dependence of the branching ratio $\calb (B \to K^{*} \gamma)$
in model II for $\tan{\beta}= 4$.
The dot-dashed and solid curve shows the central value of  the NLO model II
prediction for $\calb (B \to K^{*0} \gamma)$ and $\calb (B \to K^{*-} \gamma)$,
respectively. The region between two dashed lines shows the SM prediction:
$\calb (B \to K^{*-} \gamma) = ( 3.25^{+1.74}_{-1.43} )\times 10^{-5}$.
The shaded band shows the same data as in \fig{fig:fig1}.}
\label{fig:fig2}
\end{figure}

Fig.~\ref{fig:fig2} shows the $\mhp$ dependence of the branching ratio
$B \to K^* \gamma$ in model II for $\tan{\beta}=4$.
The dot-dashed and solid curve shows the central value of  the NLO model II
prediction for the branching ratio $\calb ( B \to \overline{K}^{*0} \gamma)$
and $\brbkm$, respectively. Other band or lines show the same thing as in
Fig.~\ref{fig:fig1}.

\begin{figure}[htb]  
\vspace{3cm}
\centerline{\mbox{\epsfxsize=14cm\epsffile{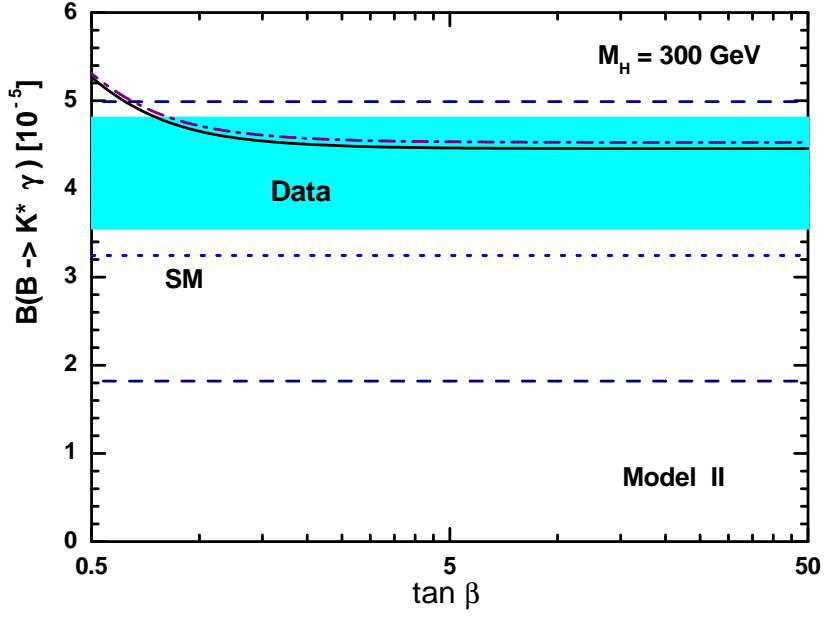}}}
\vspace{-4.5cm}
\caption{ The $\tan{\beta}$ dependence of the branching ratio $\calb (B \to K^{*} \gamma)$
in model II for $\mhp = 300$ GeV. The curves and bands have the same meaning as
in \fig{fig:fig2}. }
\label{fig:fig3}
\end{figure}

Fig.~\ref{fig:fig3} shows the $\tan{\beta}$ dependence of the branching ratio
$B \to K^* \gamma$ in model II for $\mhp=300$ GeV.
The dot-dashed and solid curve shows the central value of  the NLO model II
prediction for the branching ratio $\calb ( B \to \overline{K}^{*0} \gamma)$
and $\brbkm$, respectively. Other band or lines show the same thing as in
Fig.~\ref{fig:fig1}.

\begin{figure}[htb]  
\vspace{3cm}
\centerline{\mbox{\epsfxsize=14cm\epsffile{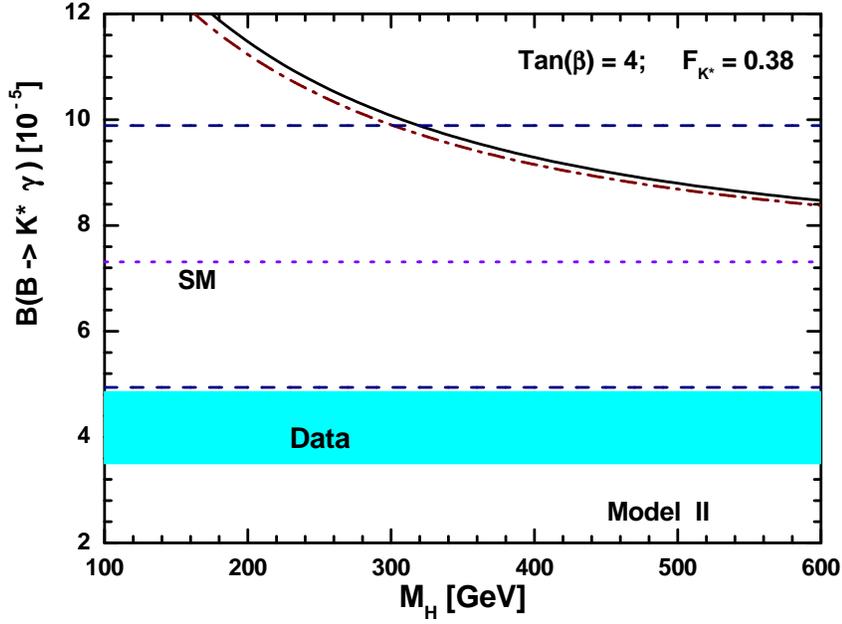}}}
\vspace{-4cm}
\caption{ The same as \fig{fig:fig2}, but for $F_{K^*}=0.38 \pm 0.06$ instead of
$F_{K^*}=0.25 \pm 0.06$.}
\label{fig:fig4}
\end{figure}

It is easy to see from Fig.\ref{fig:fig2} that a charged Higgs boson with a
mass around 200 GeV is still allowed by the measured branching ratio of
the exclusive $B \to K^* \gamma$ decay, which is weaker than the lower bound of
$\mhp \gtrsim 300$ GeV obtained from the data of the inclusive $B \to X_s \gamma$ decay.
This is consistent with general expectation. The key point here is the large uncertainty
of the non-perturbative form factor $F_{K^*}$. If we use  $F_{K^*}=0.38 \pm
0.06$ and keep all other input parameters unchanged, we get a much strong lower
limit on $\mhp$, as can be seen from Fig.\ref{fig:fig4}, where the solid and dot-dashed
curves show the NLO model II prediction for $\calb(B \to \overline{K}^{*0}\gamma)$
and $\brbkm$, respectively.
The dots line and the band between two dashed lines show the corresponding SM prediction
of $\brbkm = (7.31 ^{+2.58}_{-2.37})$ for $F_{K^*} =0.38 \pm 0.06$.

Now we study the model III. According to previous studies in
Ref.\cite{xiao03}, we got to know that the charged Higgs penguins can provide a
significant contribution to the dominant Wilson coefficient $C_7(\mu)$ and changed
its sign from negative to positive. Of course, the size of the new physics
contributions is strongly constrained by the measured branching ratio of the
inclusive $B \to X_s \gamma$, as investigated in detail in \cite{xiao03}.

For the model III-A, i.e. $(\ltt, \lbb )=(0.5, 1)$, the new physics
contributions are small,  the numerical results for $a_7^{p}(K^*\gamma)$
at the low energy scale $\mu = m_b$ are,
\beq
a_7^{u}(K^* \gamma)^{\rm III-A}&=&\underbrace{-0.3100}_{C_{7,\smallsm}(m_b)}
 +\underbrace{0.0299}_{\Delta C_{7,\smallnp} }
 \; \underbrace{-0.1394-0.0664i}_{T^I-contribution}
 \; \underbrace{+0.0336-0.0002i}_{T^{II}-contribution}\non
                    &=&-0.3859-0.0666 i, \label{eq:a7ukmb-m3a}\\
 a_7^c(K^*\gamma)^{\rm III-A}&=&\underbrace{-0.3100}_{C_{7,\smallsm}(m_b) }
 +\underbrace{0.0299}_{\Delta C_{7,\smallnp} }
 \; \underbrace{-0.0788-0.0112i}_{T^I-contribution}
 \; \underbrace{-0.0155-0.0120 i}_{T^{II}-contribution}\non
                 &=&-0.3744-0.0231 i,\label{eq:a7ckmb-m3a}
\eeq
for $\mhp =300$ GeV. The total new physics contribution
to $a_7^{p}$ in model III-A is also around $10\%$ in magnitude
for  $\mhp =300$ GeV, but in the opposite direction of that in model II.
The annihilation parts also remain unchanged.

\begin{figure}[htb]  
\vspace{3cm}
\centerline{\mbox{\epsfxsize=14cm\epsffile{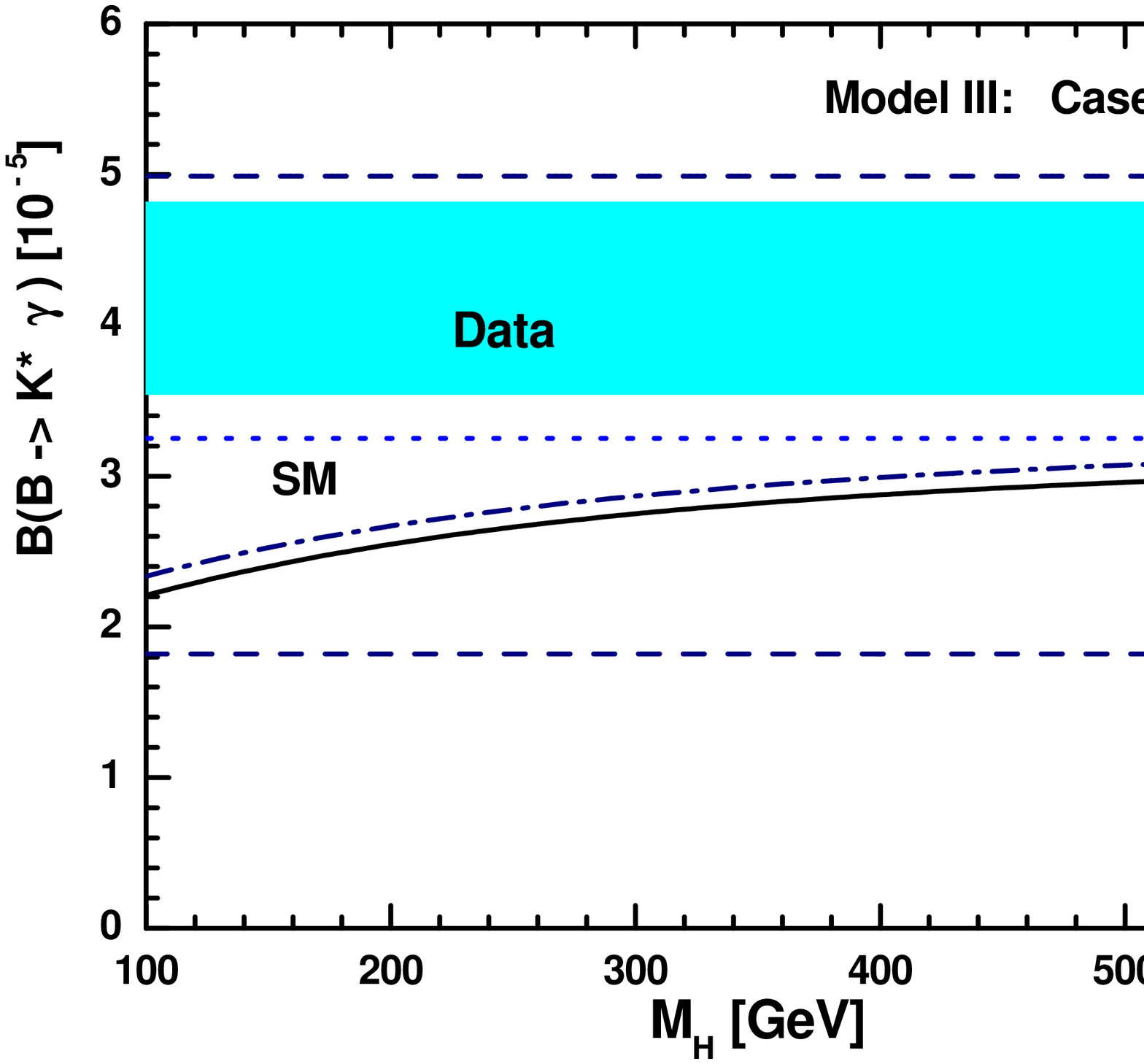}}}
\vspace{-4cm}
\caption{The $\mhp$ dependence of the branching ratio $\calb (B \to K^{*} \gamma)$
in model III-A.
The dot-dashed and solid curve shows the central value of  the NLO model III-A
prediction for $\calb (B \to \overline{K}^{*0} \gamma)$ and $\calb (B \to K^{*-} \gamma)$,
respectively. The region between two dashed lines shows the SM prediction:
$\calb (B \to K^{*-} \gamma) = ( 3.25^{+1.74}_{-1.43} )\times 10^{-5}$.
The shaded band shows the measured $\brbkm$ within $2\sigma$ errors.}
\label{fig:fig5}
\end{figure}

Fig.~\ref{fig:fig5}  shows the $\mhp$ dependence of the branching ratio
$B \to K^* \gamma$ in model III-A.
The dot-dashed and solid curve shows the central value of  the NLO model III
prediction for the branching ratio $\calb ( B \to \overline{K}^{*0} \gamma)$
and $\brbkm$, respectively. Other band or lines show the same thing as in
Fig.~\ref{fig:fig1}. The new physics contribution here is small are consistent with
the SM prediction within $1\sigma$ error. Numerically, we have
\beq
\calb (B \to \overline{K}^{*0}\gamma )^{\rm III-A}
&=& \left ( 2.87 ^{+1.50}_{-1.27}  \right ) \times 10^{-5}, \label{eq:brk0-m3a}\\
\calb (B \to            K^{*-}\gamma )^{\rm III-A}
&=& \left ( 2.75 ^{+1.48}_{-1.22}  \right ) \times 10^{-5}, \label{eq:brkp-m3a}
\eeq
for $\mhp =300$ GeV, where the four major errors as in Eqs.(\ref{eq:brk0-m2}) and
(\ref{eq:brkp-m2}) have been added in quadrature.

For the model III-B, i.e. $(\ltt, \lbb )=(0.5, 22)$, the new physics
contributions are large,  the numerical results for $a_7^{p}(K^*\gamma)$
at the low energy scale $\mu = m_b$ are,
\beq
a_7^{u}(K^* \gamma)^{\rm III-B}&=&\underbrace{-0.3100}_{C_{7,\smallsm}(m_b)}
 +\underbrace{ 0.8485}_{\Delta C_{7,\smallnp} }
 \; \underbrace{-0.1049-0.0174 i}_{T^I-contribution}
 \; \underbrace{+ 0.0752-0.0003i}_{T^{II}-contribution}\non
                    &=&0.5088-0.0177 i, \label{eq:a7ukmb-m3b}\\
 a_7^c(K^*\gamma)^{\rm III-B}&=&\underbrace{-0.3100}_{C_{7,\smallsm}(m_b) }
 +\underbrace{  0.8485}_{\Delta C_{7,\smallnp} }
 \; \underbrace{-0.0443+0.0379i}_{T^I-contribution}
 \; \underbrace{+0.0005-0.0182i}_{T^{II}-contribution}\non
                 &=& 0.4947+ 0.0196 i,\label{eq:a7ckmb-m3b}
\eeq
for $\mhp =250$ GeV. The second terms in above two equations are the new physics
corrections to the NLO Wilson coefficient $C_{7,\smallsm}(m_b)$, which is large
and positive and makes the $a_7^p(K^* \gamma)$ to be positive also.
The hard vertex and hard spectator contributions are also changed
moderately, but has only small effects on the branching ratios.

For model III-B, the theoretical predictions for branching ratios are
\beq
\calb (B \to \overline{K}^{*0}\gamma )^{\rm III-B} &=& \left [ 4.23^{+2.34}_{-1.83}(F_{K^*})
^{+0.57}_{-0.37} (\mu) ^{+0.05}_{-0.02}(\lambda_B) \pm 0.21 (m_c)\right ] \times 10 ^{-5} \non
&=& \left ( 4.23 ^{+2.42}_{-1.88}  \right ) \times 10^{-5}, \label{eq:brk0-m3b}\\
\calb (B \to            K^{*-}\gamma )^{\rm III-B} &=&\left [ 5.07 ^{+2.66}_{-2.11}(F_{K^*})
^{+0.44}_{-0.06})(\mu) ^{+0.02}_{-0.05}(\lambda_B) \pm 0.24(m_c)\right ] \times 10 ^{-5} \non
&=& \left ( 5.07 ^{+2.71}_{-2.13}  \right ) \times 10^{-5}, \label{eq:brkp-m3b}
\eeq
for $\mhp =250$ GeV.

\begin{figure}[htb] 
\vspace{3cm}
\centerline{\mbox{\epsfxsize=14cm\epsffile{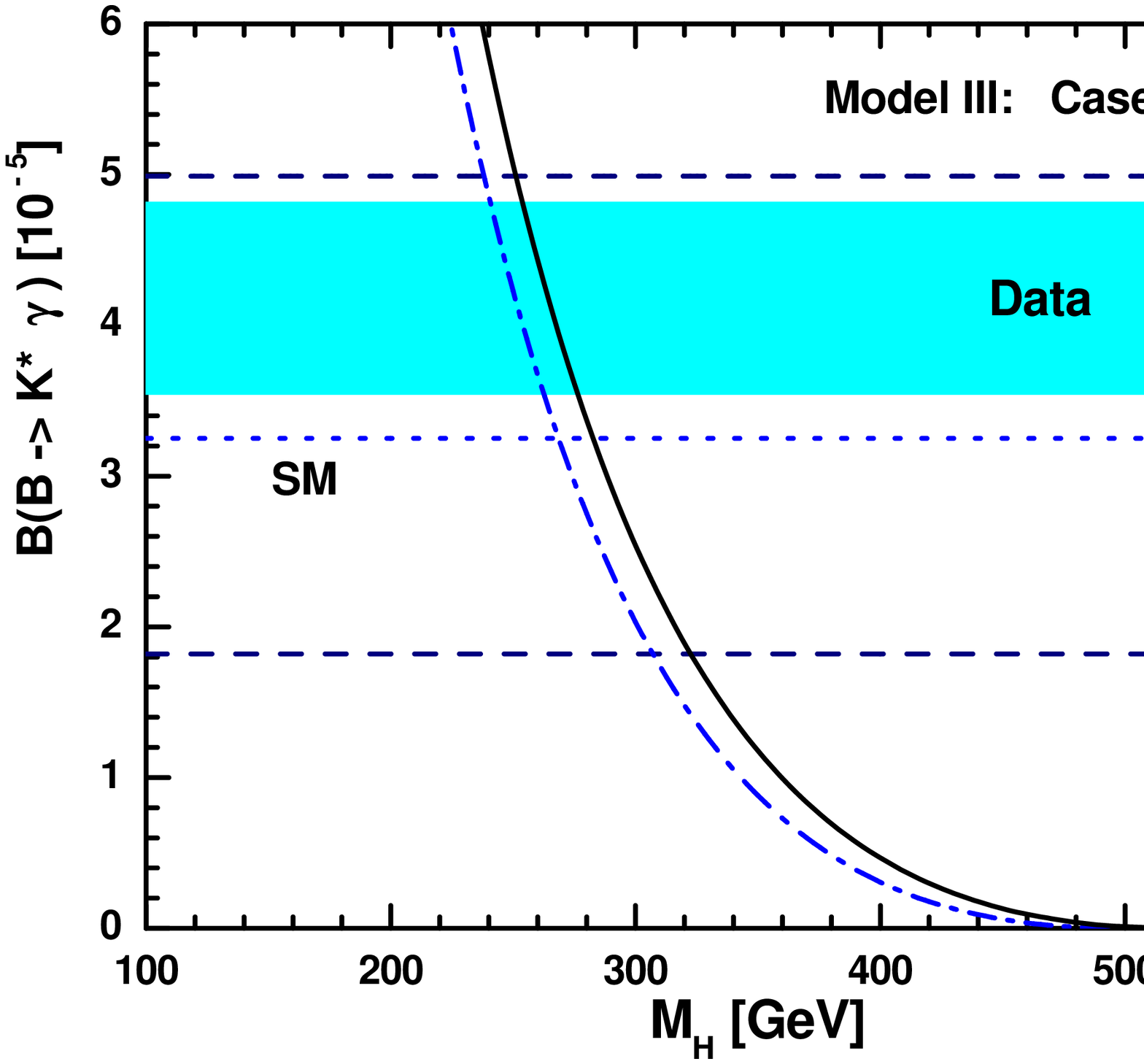}}}
\vspace{-4cm}
\caption{The same as \fig{fig:fig5}, but for model III-B, i.e.
$(\ltt, \lbb)=(0.5,22)$.}
\label{fig:fig6}
\end{figure}

Fig.~\ref{fig:fig6}  shows the $\mhp$ dependence of the branching ratio
$B \to K^* \gamma$ in model III-B.
The dot-dashed and solid curve shows the central value of  the NLO model III-B
prediction for the branching ratio $\calb ( B \to \overline{K}^{*0} \gamma)$
and $\brbkm$, respectively. Other band or lines show the same thing as in
Fig.~\ref{fig:fig5}.

If we add  the theoretical errors as given in Eqs.(\ref{eq:brk0-m3b},\ref{eq:brkp-m3b})
with the corresponding experimental errors
in Eqs.(\ref{eq:bvga-exp}) in quadrature and treat them as the total $1\sigma$ error,
we then read off the allowed regions of  $\mhp$ from \fig{fig:fig6},
\beq
218 \leq \mhp \leq 293 {\rm GeV}, \ \ {\rm and} \ \ \mhp \geq 1670 {\rm GeV},
\label{eq:lka}
\eeq
allowed by the measured $\calb ( B \to \overline{K}^{*0} \gamma)$, and
\beq
226 \leq \mhp \leq 315 {\rm GeV}, \ \ {\rm and} \ \ \mhp \geq 1490 {\rm GeV},
\label{eq:lkb}
\eeq
allowed by the measured $\calb ( B \to K^{*-} \gamma)$. These constarints
on $\mhp$ are well consistent with those
obtained from the inclusive $B \to X_s \gamma$ decays as given in Ref.\cite{xiao03}.
Of course, the large theoretical error is dominated by the  uncertainty of the
form factor $F_{K^*}$ here.

For the exclusive $B \to K^* \gamma$ decay, the theoretical prediction for the
CP symmetry $\acp$ as defined in Eq.(\ref{eq:acp-vks}) is very small:
\beq
\left |\acp(B \to K^* \gamma) \right |  < 1\%
\eeq
in the SM and all three types of the 2HDM's considered here, which is consistent with
the measurements  as reported by BaBar \cite{babar-vg}  and Belle
Collaboration \cite{belle-vg}:
\beq
\acp(B \to K^* \gamma) &=& [-0.17, +0.082],  \\
\acp(B \to K^* \gamma) &=& -0.001 \pm 0.044 \pm 0.008.
\eeq

\subsection{Isospin Symmetry}

As can be seen in last subsection, the large uncertainty of the form factor $F_{K^*}$
dominates the total error of the theoretical prediction of the branching ratios.
For the isospin symmetry breaking of $B \to K^* \gamma$ system, however,
its dependence on the form factor  $F_{K^*}$ largely cancelled in the ratio.
From Eqs.(\ref{eq:br-vga},\ref{eq:rv}), the isospin symmetry
breaking $\Delta_{0-}(K^*\gamma)$ as
defined in Eq.(\ref{eq:iso}) can also be written as
\beq
\Delta_{0-}(K^* \gamma) = \frac{ \eta_{\tau} |R_{\overline{K}^{*0}}|^2 -
|R_{K^{*-}} |^2 }{\eta_{\tau} |R_{\overline{K}^{*0}}|^2 + |R_{K^{*0}}|^2  }
\label{eq:isob}
\eeq
where $R_{K^*}$ have been given in Eq.(\ref{eq:rks}). In our approximation,
the isospin breaking is generated by weak annihilation
contributions, and has a residue sensitivity to the form factors $F_V$
induced through the $F_V$ dependence of $b^V$ and $d^V$ functions as defined in
Eqs.(\ref{eq:bv},\ref{eq:dv}).
Since $\lambda^s_{u}= V_{us}^* V_{ub}$ is about two orders smaller than
$\lambda^s_c = V_{cs}^* V_{cb}$, the function $R_{K^*}$ is largely determined
by $a_7^c(K^*\gamma)$.

In the SM,  we have numerically
\beq
\Delta_{0-}(K^* \gamma)^{\rm SM} &=&  \left [ 5.6^{+1.7}_{-1.1}(F_{K^*})
^{+4.0}_{-2.1} (\mu) ^{+0.6}_{-1.4}(\lambda_B) \pm 0.1 (m_c)\right ] \times 10 ^{-2} \non
&=& \left ( 5.6 ^{+4.4}_{-2.8}  \right ) \times 10^{-2}, \label{eq:d0m-sm}
\eeq
where the errors added in quadrature. The dominant error comes from the uncertainty of the low
energy scale $1/2 m_b \leq \mu \leq 2 m_b$. The SM prediction agrees well with
the measured value of $\Delta_{0-}^{\rm exp}(K^* \gamma) = (3.9\pm 4.8)\% $.

In the general two-Higgs-doublet models, by assuming $\tan{\beta}=4$ and $\mhp=250$ GeV,
we find numerically
\beq
\Delta_{0-}(K^* \gamma)^{\rm I} &=&  \left ( + 5.7^{+4.3}_{-2.7} \right )
\times 10^{-2}, \label{eq:d0m-m1}\\
\Delta_{0-}(K^* \gamma)^{\rm II} &=&  \left ( + 4.6^{+3.7}_{-2.4} \right )
\times 10^{-2}, \label{eq:d0m-m2}\\
\Delta_{0-}(K^* \gamma)^{\rm III-A} &=&  \left ( + 6.2^{+4.7}_{-3.0} \right )
\times 10^{-2}, \label{eq:d0m-m3a}\\
\Delta_{0-}(K^* \gamma)^{\rm III-B} &=&  \left ( -5.1^{+2.6}_{-4.3} \right )
\times 10^{-2}, \label{eq:d0m-m3b}
\eeq
where the errors induced by the uncertainties of $\mu, F_{K^*}, \lambda_B$ and $m_c$ have
been added in quadrature, and the uncertainty of $\mu$ dominates the total
theoretical error.

\begin{figure}[htb]  
\vspace{3cm}
\centerline{\mbox{\epsfxsize=14cm\epsffile{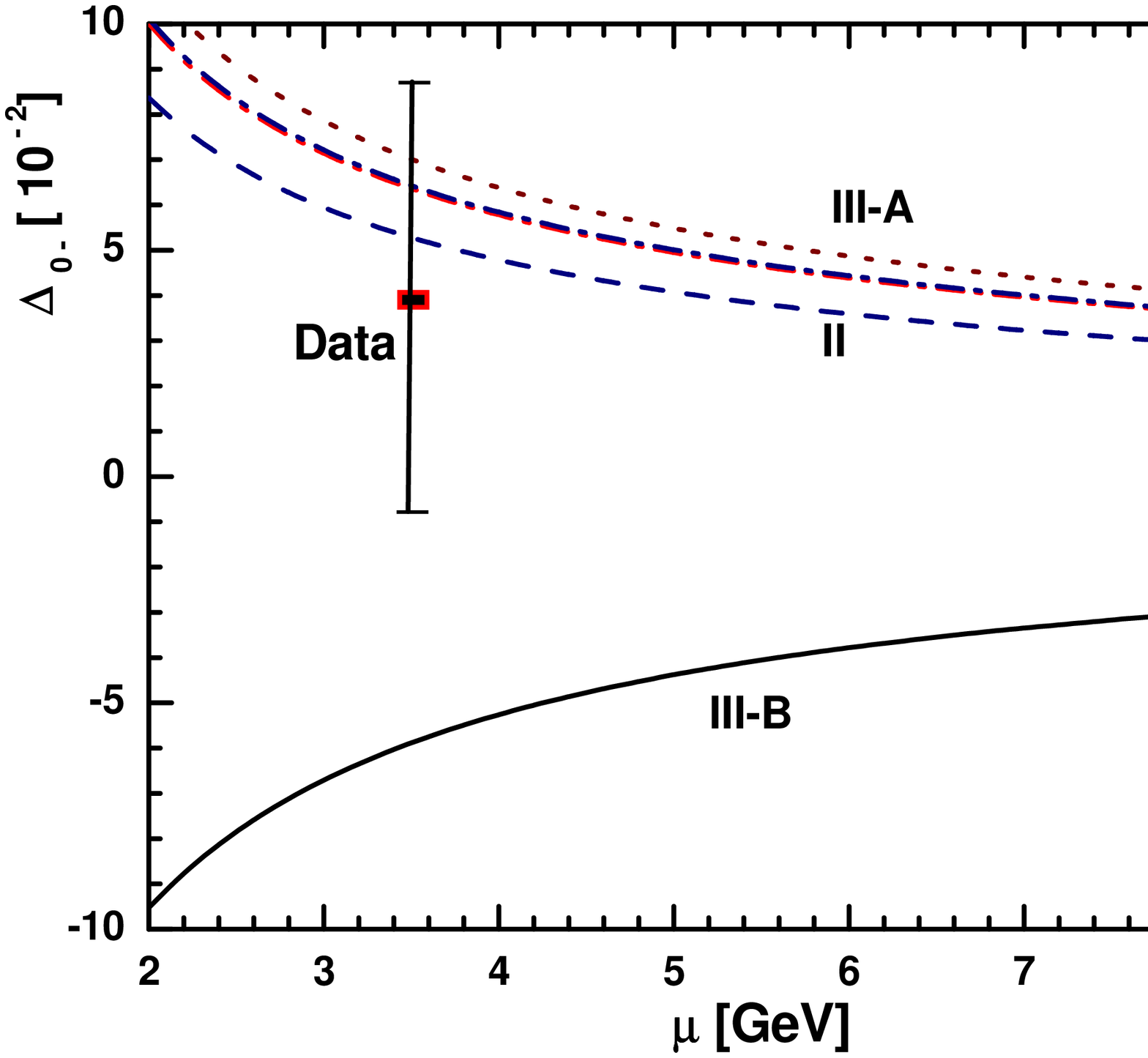}}}
\vspace{-4cm}
\caption{The $\mu$ dependence of the isospin symmetry breaking
$\Delta_{0-}(K^*\gamma)$ in the SM
and the general 2HDM's. The error bar shows the data. For details see the text.}
\label{fig:fig7}
\end{figure}

Fig.~\ref{fig:fig7}  shows the $\mu$ dependence of the isospin symmetry breaking
$\Delta_{0-}(K^* \gamma)$ in the general 2HDM's for $\tan{\beta}=4$ and $\mhp=250$ GeV.
Two coincided dot-dashed curves show the SM and model I prediction ,
the dash and dots curve show the model II and model III-A prediction  respectively,
and the solid curve refers to the model III-B prediction.
The error bar shows the data $\Delta_{0-}(K^* \gamma)^{\rm exp}= (3.9 \pm 4.8)\%$.

\begin{figure}[htb] 
\vspace{3cm}
\centerline{\mbox{\epsfxsize=14cm\epsffile{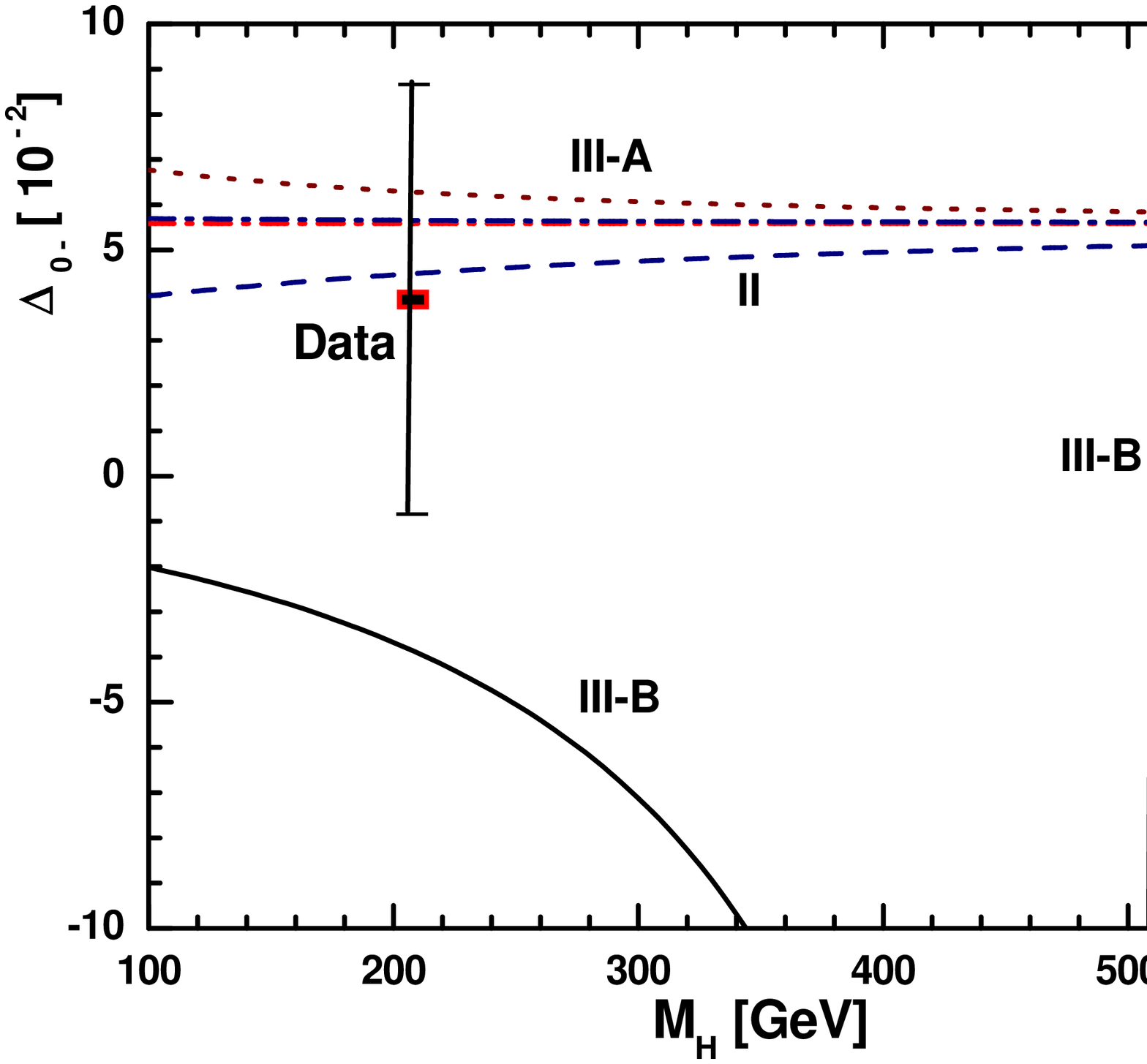}}}
\vspace{-4cm}
\caption{The $\mhp$ dependence of the isospin symmetry breaking
$\Delta_{0-}(K^*\gamma)$ in the SM
and the general 2HDM's. The error bar shows the data. For details see the text.}
\label{fig:fig8}
\end{figure}

Fig.~\ref{fig:fig8}  shows the $\mhp$ dependence of the isospin symmetry breaking
$\Delta_{0-}(K^* \gamma)$ in the general 2HDM's for $\tan{\beta}=4$ and $\mu=m_b$ GeV.
Two coincided dot-dashed curves show the SM and model I predictions.
The dashed, dots solid curve show the model II, III-A and III-B prediction, respectively.
The error bar shows the data as in \fig{fig:fig7}.

\begin{figure}[htb] 
\vspace{2cm}
\centerline{\mbox{\epsfxsize=14cm\epsffile{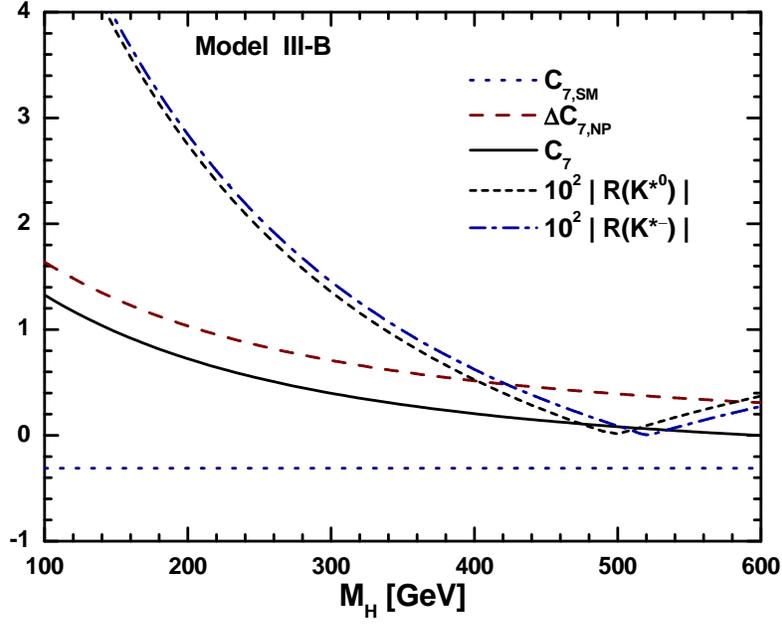}}}
\vspace{-4.5cm}
\caption{Plots of the $\mhp$ dependence of $C_{7,\smallsm}(m_b)$ (horizontal dots line),
$\Delta C_{7,\smallnp}$ (dashed curve), $C_7(m_b)$ (solid curve),
$10^2 |R(\overline{K}^{*0}\gamma)|$ (short-dashed curve) and
$10^2 |R(K^{*-}\gamma)|$(dot-dashed curve).}
\label{fig:fig9}
\end{figure}

From above two figures, one can see that only the theoretical prediction of the
model III-B is rather different from that of the SM and looks like deviating from
the data. But the regions $\mhp < 200$ GeV and $300 \lesssim
\mhp \lesssim 1500$ GeV have  been excluded by the data of inclusive $B \to X_s \gamma$
\cite{xiao03} and by the constraint as illustrated in \fig{fig:fig6}.
The main reason for the great changes of the solid curve in \fig{fig:fig8} is
the strong cancellation between the negative $C_{7,\smallsm}(m_b)$ and its positive
new physics counterpart as illustrated clearly in \fig{fig:fig9}, where the
solid curve shows the summation of the SM and new physics contributions to the
dominant Wilson coefficient $C_7$, i.e.,
$C_7(m_b) = C_{7,\smallsm}(m_b) + \Delta C_{7,\smallnp}(m_b)$. When $C_7(m_b)$
approaches zero, the summation of other ``originally small" parts (such as the $T_i^I$
, $T_i^{II}$ and $a_{ann}^p$ contributions ) becomes important and leads to
an abnormally large isospin breaking.  The short-dashed and dot-dashed curves
in \fig{fig:fig9} shows the absolute value of $R_{\overline{K}^{*0}}$
and $R_{K^{*-}}$,
respectively. The isospin breaking is proportional to the difference of their
squares. At the close region of the crossing point of $R_{\overline{K}^{*0}}$ and
$R_{K^{*-}}$, the ratio $\Delta_{0-}(K^* \gamma)$ can be large and changes
the sign. But as mentioned previously, this region around $\mhp=500$ GeV has been
excluded by the data of branching ratios from both the inclusive and exclusive radiative
B meson decays.

In the region of $\mhp \sim 250 GeV$, the model III-B is disfavered by the measured value
of isospin breaking $\Delta_{0-}(K^*\gamma)$ as can be seen from Figs.~\ref{fig:fig7} and
\ref{fig:fig8}. But taking the sizable experimental and theoretical uncertainties into account,
the theoretical prediction of the model III-B is still compatible with
the data within $2\sigma$ errors.
In another word, the positive $C_7(m_b)$ is disfavored but can not
be excluded by the present data.

\section{$B \to \rho \gamma$} \label{sec:brho}

When compared with $B \to K^* \gamma$ decay, the $B \to \rho \gamma$ decay mode
is particularly interesting in search for new physics beyond the SM, because of the
suppression of $b \to d$ transitions in the SM and the simultaneous chirality suppression.
For $B \to \rho \gamma$ decay, we generally know that:
(a) both $a_7^u$ and $a_7^c$ contribute
effectively since $\lambda_u^d$ and $\lambda_c^d$ are comparable in magnitude;
(b) the branching ratios of $B \to \rho \gamma$ are suppressed with respect to $B \to K^* \gamma$
by roughly a factor  of $|V_{td}/V_{ts}|^2 \approx 4 \times 10^{-2}$;
(c) the CP asymmetry $\acp(B \to \rho \gamma)$ is generally at $10\%$ level, and
may be observed in B factory experiments;
(d) the new physics may provide significant contribution to the observables of
$B \to \rho \gamma$ decay;
and (e) only the experimental upper limits on the branching ratios of
$B \to \rho \gamma$ are available now.

\subsection{Branching ratios and CP asymmetries}

From Eq.(\ref{eq:br-vga}), the branching ratios of $B \to \rho \gamma$ decays
can be written as
\beq
\calb (\overline{B}\to \rho \gamma)& =& \tau_B\frac{G_F^2\alpha m_B^3m_b^2}{32\pi^4}
\left (1-\frac{m_\rho^2}{m_B^2} \right )^3
\left | R_{\rho} \right |^2 c_\rho^2  |F_\rho|^2,\label{eq:br-rho}
\eeq
with
\beq
R_{\rho} =   V_{ud}^* V_{ub} \,\left [  a_7^u(\rho \gamma) + a_{ann}^u (\rho \gamma) \right ]
+  V_{cd}^* V_{cb}\,\left [  a_7^c(\rho \gamma) +a_{ann}^c (\rho \gamma) \right ].
\label{eq:rrho}
\eeq

By using the formulas  as given in Eqs.(\ref{eq:a7vga}) and (\ref{eq:annrho}) and the
central values of the input parameters in \tab{input}, we find the SM predictions for
$a_7^{p}(\rho \gamma)$ and $a_{ann}(\rho\gamma)$ at the low energy scale $\mu = m_b$,
\beq
a_7^{u}(\rho \gamma)&=&\underbrace{-0.3212}_{C^0_{7,\smallsm}(m_b)}
 +\underbrace{0.0113}_{\Delta C^1_{7,\smallsm} }
 \; \underbrace{-0.1407-0.0683i}_{T^I-contribution}
 \; \underbrace{0.0343-0.0003i}_{T^{II}-contribution}\non
                    &=&-0.4164-0.0686i, \label{eq:a7urmb}\\
 a_7^c(\rho \gamma)&=&\underbrace{-0.3212}_{C^0_{7,\smallsm}(m_b) }
 +\underbrace{0.0113}_{\Delta C^1_{7,\smallsm} }
 \; \underbrace{-0.0802-0.0131i}_{T^I-contribution}
 \; \underbrace{-0.0166-0.0143i}_{T^{II}-contribution}\non
                 &=&-0.4067-0.0274i,\label{eq:a7crmb}\\
a^u_{ann}(\rho^0\gamma) &=&-0.0032, \ \  a^c_{ann}(\rho^0\gamma) =-0.0127, \non
a^u_{ann}(          \rho^- \gamma)  &=& 0.1883, \ \ a^c_{ann}(    \rho^- \gamma)  = 0.0032.
\label{eq:ann-rho}
\eeq
Here the values of weak annihilation factors are slightly different from those for
$B \to K^* \gamma$ decay, and only the $T^{II}$ contributions to $a^p_7(\rho \gamma)$ are
different from those to $a_7^p(K^* \gamma)$ because of the small differences of the
$H_i$ functions between two decay modes as can be seen in Appendix A.

The corresponding NLO SM predictions for branching ratio $\calb (B \to \rho \gamma)$ are
\beq
\calb (B \to \rho^0 \gamma )^{SM} &=& \left [ 0.91 \pm 0.29 (\gamma) ^{+0.25}_{-0.22}(F_{\rho})
\pm 0.17 (\mu) ^{+0.10}_{-0.03}(\lambda_B) \pm 0.10 (m_c)\right ] \times 10 ^{-6} \non
&=& \left ( 0.91 ^{+0.42}_{-0.40}  \right ) \times 10^{-6}, \label{eq:brrho0-sm}\\
\calb (B \to \rho^- \gamma )^{SM} &=&\left [ 2.03 \pm 0.34 (\gamma) ^{+0.54}_{-0.47} (F_\rho)
\pm 0.31(\mu) ^{+0.46}_{-0.13}(\lambda_B)\pm 0.12 (m_c) \right ] \times 10 ^{-6} \non
&=& \left ( 2.03 ^{+0.85}_{-0.67}  \right ) \times 10^{-6}, \label{eq:brrhom-sm}
\eeq
where the individual errors have been added in quadrature. The uncertainties
of the CKM angle $\gamma$ ( here we take $ \gamma = (60 \pm 20)^\circ$ in the calculation)
and the form factor $F_\rho$ dominate the total error, and the remaining errors
from other input parameters are negligibly small.

\begin{figure}[htb]  
\vspace{3cm}
\centerline{\mbox{\epsfxsize=14cm\epsffile{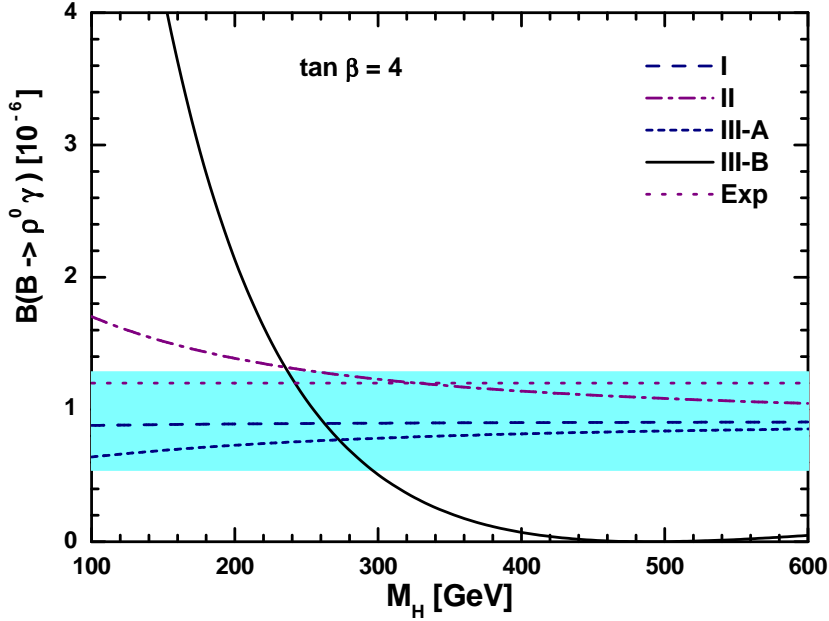}}}
\vspace{-4cm}
\caption{Plots of the $\mhp$ dependence of the branching ratio $B \to \rho^0 \gamma$
in the SM (shaded band), the model I and II (dashed and dot-dashed curves), and the
model III-A (short-dashed curve) and III-B (solid curve). The horizontal
dots line shows the experimental upper bound (at $9\%$C.L. \cite{babar-vg}):
$\calb(B \to \rho^0\gamma) < 1.2\times 10^{-6}$}
\label{fig:fig10}
\end{figure}

\begin{figure}[htb] 
\vspace{3cm}
\centerline{\mbox{\epsfxsize=14cm\epsffile{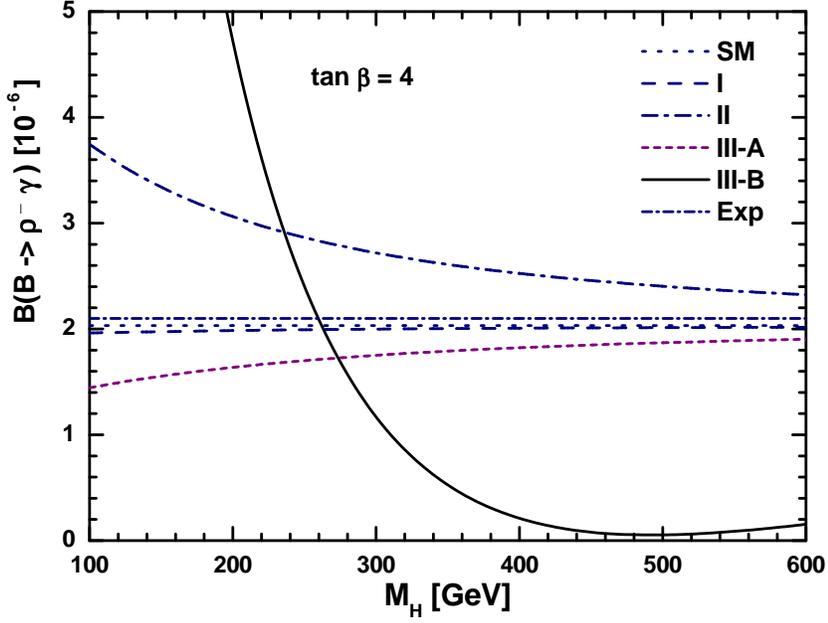}}}
\vspace{-4cm}
\caption{The same as \fig{fig:fig10} but for the decay $B \to \rho^- \gamma$.}
\label{fig:fig11}
\end{figure}

The central values and theoretical uncertainties of the branching ratios
$\calb (B \to \rho \gamma)$ in the SM and the general 2HDM's are all listed in
Table \ref{rho}. The SM prediction is
well consistent with the experimental upper limits within $1\sigma$ error.
The predictions of the model I , II and III-A are also compatible with the data
within errors, as illustrated in
\fig{fig:fig10} and \fig{fig:fig11} for $B \to \rho^0 \gamma$ and $B \to \rho^-\gamma$,
respectively. For model III-B, however, the branching ratios can be changed
significantly when the charged Higgs boson is light or heavy, as illustrated by the solid curves
in \fig{fig:fig10} and \fig{fig:fig11}. From the experimental upper bounds
on  $\calb (B \to \rho \gamma)$ as given in Table \ref{bvga}, we find the
lower limit on $\mhp$
\beq
\mhp \geq 206 {\rm GeV}
\label{eq:lra}
\eeq
when  the $2\sigma$ theoretical errors are also taken into account. This
lower bound is compatible with those obtained  from the measured $\calb (B \to K^* \gamma)$
as given in Eqs.(\ref{eq:lka}) and (\ref{eq:lkb}) and from the inclusive $B \to X_s \gamma$
decay \cite{xiao03}.

\begin{figure}[htb] 
\vspace{3cm}
\centerline{\mbox{\epsfxsize=14cm\epsffile{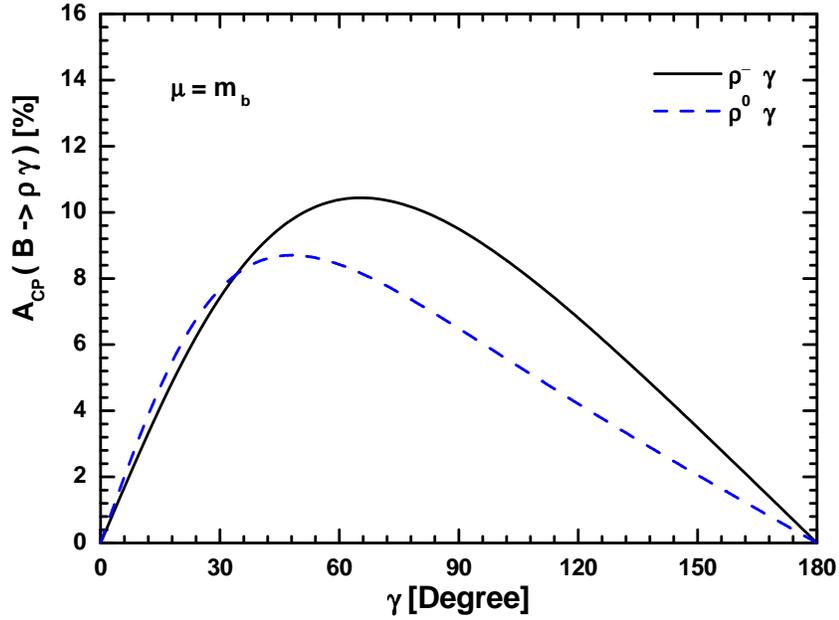}}}
\vspace{-4.5cm}
\caption{Plots of the angle $\gamma$ dependence of the CP asymmetries for
 $B \to \rho^0 \gamma$ (dashed curve) and $B^\pm \to \rho^\pm \gamma$ (solid curve)
 decays in the SM.}
\label{fig:fig12}
\end{figure}

The CP asymmetry of $B \to \rho \gamma$ decays is defined in the same way as for
$B \to K^*\gamma$ decays in Eq.(\ref{eq:acp-vks}).
Using the input parameters as listed in Table \ref{input}, one
finds the NLO SM predictions,
\beq
\acp (\rho^0 \gamma )^{SM} &=& \left [ 8.4 ^{+3.8}_{-1.8} (\mu)\pm 1.9 (R_b)
\pm 0.8 (\lambda_B) ^{+0.9}_{-1.1}(m_c) \pm 0.4 (F_\rho)
^{+0.1}_{-1.2}(\gamma) \right ] \times 10^{-2}\non
&=& \left ( 8.4 ^{+4.4}_{-3.2}  \right ) \times 10^{-2}, \label{eq:acprho1-sm}\\
\acp (\rho^\pm \gamma )^{SM} &=& \left [ 10.4 ^{+5.4}_{-2.5} (\mu)\pm 2.4 (R_b)
^{+0.3}_{-0.0}(\lambda_B) \pm 0.8(m_c) \pm 0.1 (F_\rho)
^{-0.3}_{-1.4}(\gamma) \right ] \times 10^{-2}\non
&=& \left ( 10.4 ^{+6.0}_{-3.8}  \right ) \times 10^{-2}, \label{eq:acprho2-sm}
\eeq
where the errors have been added in quadrature. The CP asymmetry of $B \to \rho \gamma$ is
large is size and depends sensitively on the variations of the scale $\mu$ and
$R_b=\sqrt{\bar{\rho}^2 + \bar{\eta}^2}$.
If we consider the whole range of $0^\circ \leq \gamma \leq 180^\circ$ instead of
$\gamma = (60\pm 20)^\circ$ preferred by the global fit result \cite{pdg2003}
, the CP asymmetry $\acp(B \to \rho \gamma)$ also shows a strong dependence
on the angle $\gamma$ as illustrated by \fig{fig:fig12} for $B^\pm \to \rho^\pm \gamma$
(solid curve) and $B \to \rho^0 \gamma$ (dashed curve) decays.

The numerical values of CP asymmetries in the SM and the general 2HDM's are also
listed in Table \ref{rho}. The theoretical predictions of the SM and the model
I, II and III-A are all compatible, around $+ 10\%$. The CP asymmetry
$\acp (B \to \rho \gamma)$ in model III-B, however, is comparable in size with
the SM prediction, but has an opposite sign, as shown in \fig{fig:fig13},
where the upper and lower three curves show the theoretical predictions for
$\mu=m_b/2$ (dashed curves), $m_b$ (solid curves) and $2m_b$ (dots curves) in
the SM and the model III-B, respectively. For the $B \to \rho^0 \gamma$ decay
mode, we have the similar conclusion. This feature may be served as a good observable
to distinguish the model III-B (or a positive $C_7(m_b)$) with the SM (a negative
$C_7(m_b)$).

\begin{table}
\begin{center}
\caption{The NLO theoretical predictions for branching ratios and CP asymmetries
in the SM and the models I, II, III-A and III-B, assuming $\tan{\beta}=4$ and
$\mhp=250$GeV. The errors induced by the uncertainties of six
input parameters ($\mu, R_b, \lambda_B,m_c, F_\rho$ and $\gamma$)
are taken into account. Individual errors are added in quadrature. }
\label{rho}
\begin{tabular}{c| c c c c c } \hline
Decays &\ \  SM \ \ &\ \ model I\ \ &\ \ model II\ \ &\ \ model III-A\ \ &\ \ model III-B\ \  \\ \hline\hline
$\calb (\overline{B}^{0} \to \rho^0 \gamma) (10^{-6})$
& $0.91 ^{+0.42}_{-0.40}$ & $0.90^{+0.41}_{-0.39}$& $1.30^{+0.59}_{-0.51}$
& $0.76^{+0.35}_{-0.33}$& $1.07 ^{+0.50}_{-0.41}$\\ \hline
$\calb (B^{-} \to \rho^- \gamma) (10^{-6})$
& $2.0 ^{+0.8}_{-0.7} $& $2.0 ^{+0.8}_{-0.7}$& $2.9^{+1.2}_{-0.9}$& $1.7 ^{+0.7}_{-0.6}$
& $2.4 ^{+1.4}_{-1.1}$\\ \hline \hline
$\acp (\overline{B}^{0} \to \rho^0 \gamma) (\%)$
& $8.4 ^{+4.4}_{-3.2} $& $8.5^{+4.4}_{-3.5}$& $7.0^{+3.8}_{-3.0}$& $9.3^{+4.8}_{-3.9}$
& $-7.2 ^{+3.6}_{-4.6}$\\ \hline
$\acp (B^{-} \to \rho^- \gamma) (\%)$
& $10.4 ^{+6.0}_{-3.8} $& $10.5^{+6.0}_{-3.9}$& $8.7^{+5.1}_{-2.7}$& $11.4^{+6.5}_{-3.9}$
& $-8.5 ^{+4.2}_{-5.0}$\\ \hline \hline
\end{tabular}
\end{center}
\end{table}

\subsection{Isospin and U-spin symmetries }

According to currently available data, the $SU(2)$ isospin symmetry of the strong
interaction is a very good symmetry with a breaking no more than $5\%$.
The U-spin symmetry, the $SU(3)$ flavor symmetry of the strong interaction
under exchanges of the down and strange quarks, however, may have a breaking
around $20\%$ (i.e., $\sim (F_K/F_\pi -1)$ ) as frequently used in the study of
$B \to K \pi$ decays. For the exclusive $B \to K^* \gamma$ decays, the isospin
breaking derived from the measured branching ratios is indeed around $5\%$ as
given in Eq.(\ref{eq:d0m-exp}). For $B \to \rho \gamma$ decays, no measurements
are available now.

As in Refs.\cite{bosch02a,bosch02b}, we also define the isospin
symmetry breaking of $B \to \rho \gamma$ decays as the form of
\beq
\Delta (\rho\gamma) = \frac{1}{2}\left [
\frac{\Gamma(B^+ \to \rho^+ \gamma)}{2\Gamma(B^0\to \rho^0 \gamma)}
 + \frac{\Gamma(B^+ \to \rho^+ \gamma)}{2\Gamma(B^0\to \rho^0 \gamma)} -2 \right
 ]. \label{eq:iso1}
\eeq

Using the central values of input parameters as listed in Table \ref{input} and assuming
$tan{\beta}=4$, $\mhp=250$ GeV, we find numerically that
\beq
\Delta (\rho\gamma)&=& \left \{\begin{array}{ll}
\left ( 0.9 ^{+23.3}_{-13.5} \right )\times 10^{-2} & {\rm in \ \ SM}, \\
\left ( 0.9 ^{+23.3}_{-13.6} \right )\times 10^{-2} & {\rm in \ \ model \ \ I}, \\
\left ( 0.4 ^{+18.3}_{-11.1} \right )\times 10^{-2} & {\rm in \ \ model \ \ II}, \\
\left ( 1.3 ^{+25.9}_{-14.9} \right )\times 10^{-2} & {\rm in \ \ model \ \ III-A}, \\
\left ( 4.9 ^{+12.0}_{-14.6} \right )\times 10^{-2} & {\rm in \ \ model \ \ III-B}, \\
\end{array} \right. \label{eq:iso2}
\eeq
where the errors from uncertainties of input parameters have been added in quadrature.
The largest theoretical uncertainty comes from the the CKM angle $\gamma$.

In \fig{fig:fig14}, we show the angle $\gamma$ dependence of the isospin
breaking $\Delta (\rho \gamma)$ in the SM and the considered 2HDM's for
$\tan{\beta}=4$, $\mhp=250$ GeV and $0^\circ \leq \gamma \leq 180^\circ$.
It is easy to  see from \fig{fig:fig14} that
(a) except for the model III-B, the isospin breaking in the SM and other 2HDM's have
the similar $\gamma$ dependence;
(b) all theoretical predictions become almost identical and
very small in magnitude for $\gamma \sim 55^\circ$(the value preferred by the global fit results),
and the smallness of $\Delta (\rho\gamma)$
is also consistent with the general expectation and other measurements;
(c) the theoretical predictions in the SM and model III-B have a very different $\gamma$
dependence, and have the opposite sign for small or large values of the CKM
angle $\gamma$.

The U-spin symmetry is another interesting observable for $B \to (K^*, \rho) \gamma$
decays, and has been studied for example in Refs.\cite{bosch02a,bosch02b,isosm}.
In the limit of U-spin symmetry, the quantity
\beq
\Delta U(K^*,\rho)\equiv \Delta \calb(B \to K^* \gamma) + \Delta \calb(B \to \rho \gamma)
\equiv 0
\eeq
with
\beq
\Delta \calb(B \to K^* \gamma) &=& \calb (B^+ \to K^{*+} \gamma) - \calb (B^- \to K^{*-}
\gamma), \\
\Delta \calb(B \to \rho \gamma)&=& \calb (B^+ \to \rho^+ \gamma) - \calb (B^- \to \rho^-
\gamma),
\eeq
should be satisfied. Using the central values of input parameters, we find
the SM prediction of $\Delta U(K^*, \rho)$
\beq
\Delta \calb(B \to K^* \gamma) &=&-3.7\times 10^{-7}, \\
\Delta \calb(B \to \rho \gamma)&=&+4.4\times 10^{-7},
\eeq
where we have chosen $\gamma=90^\circ$ which maximizes the effects. The two parts
have opposite sign and cancels to a large extent, leaving a small U-spin
breaking
\beq
\Delta U(K^*, \rho) =0.7 \times 10^{-7}\label{eq:dsb1}.
\eeq
in the SM, which is only about $8\%$ of  $\calb (B \to \rho^0 \gamma)$.
In the general 2HDM's, we find the numerical results
\beq
\Delta U(K^*, \rho)&=& \left \{\begin{array}{ll}
0.7  \times 10^{-7} & {\rm in \ \ model \ \ I}, \\
0.9  \times 10^{-7} & {\rm in \ \ model \ \ II}, \\
0.6  \times 10^{-7} & {\rm in \ \ model \ \ III-A}, \\
-1.5\times 10^{-7} & {\rm in \ \ model \ \ III-B}, \\
\end{array} \right. \label{eq:usb1}
\eeq
for $\tan{\beta}=4$, $\mhp=250$ GeV and $\gamma=90^\circ$. The new physics
contributions in the conventional model I , II and the model III-A
have little effect on the size of U-spin symmetry breaking. In model III-B,
although $\Delta U(K^*, \rho)$ becomes negative, but it is still small in magnitude.

\section{Conclusions}

By employing the QCD factorization approach for the exclusive $B \to V \gamma$ decays
as proposed in Refs.\cite{beneke01,ali02a,bosch02a}, we calculated the NLO new
physics contributions to the branching ratios,
CP asymmetries, isospin symmetry breaking and U-spin symmetry breaking of the
exclusive radiative decays
$B \to K^*\gamma$ and $B \to \rho \gamma$, induced by the charged Higgs penguin diagrams
appeared in the general two-Higgs-doublet models including the conventional model I and II,
as well as two-typical cases of model III.
The NLO new physics contributions
are included through their  corrections to the NLO Wilson coefficients $C_7(\mw)$ and
$C_8(\mw)$ at the matching scale $\mw$.

In section \ref{sec:th}, we gave a brief review about the effective Hamiltonian
and the calculation of the exclusive $B \to V \gamma$ ($V=K^*, \rho$) decays at
next-to-leading order in QCD factorization, presented the relevant formulas
for the calculation of Wilson coefficients and physical observables in the SM and the
general two-Higgs-doublet models.

\begin{figure}[tb]  
\vspace{-1cm}
\centerline{\mbox{\epsfxsize=14cm\epsffile{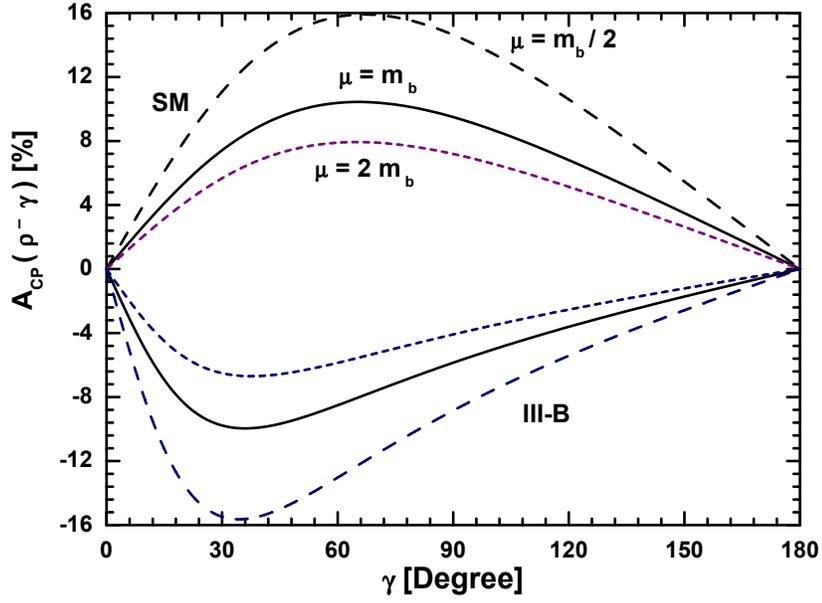}}}
\vspace{-1cm}
\caption{Plots of the angle $\gamma$ dependence of the CP asymmetry for
 $B^\pm \to \rho^\pm  \gamma$ decay in the SM and model III-B for $\mu=m_b/2$
 (dots curves), $m_b$ (solid curves ) and $2m_b$ (dashed curves).}
\label{fig:fig13}
\end{figure}

\begin{figure}[tb]  
\vspace{3cm}
\centerline{\mbox{\epsfxsize=14cm\epsffile{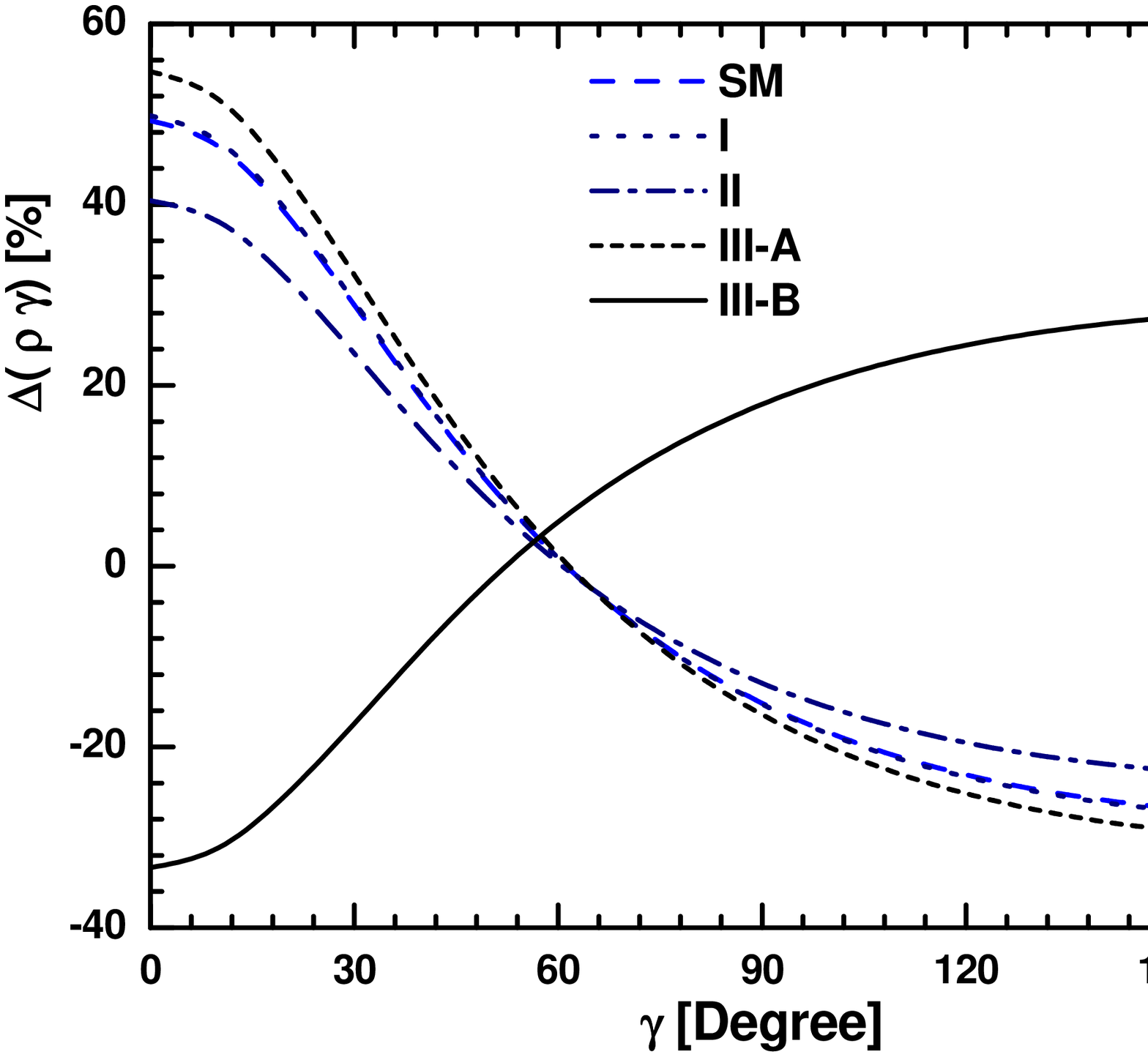}}}
\vspace{-4cm}
\caption{The isospin breaking $\Delta(\rho \gamma)$ vs the CKM angle $\gamma$
in the SM and general 2HDM's for $\tan{\beta}=4$ and $\mhp=250$ GeV.}
\label{fig:fig14}
\end{figure}

In section \ref{sec:bks} and \ref{sec:brho}, we calculated the NLO new physics
contributions to the branching ratios and other observables of
$B \to K^* \gamma$ and $B \to \rho \gamma$ decays in the
general 2HDM's, compared the theoretical predictions with those currently
available experimental measurements, and found the following points:

\begin{itemize}

\item
The new physics corrections to  the physical observables under consideration in
this paper
are generally small in the model I and model III-A, moderate in model II, but large
in model III-B.  And therefore the theoretical predictions in the SM, model I and III-A
are always in good agreement with the corresponding data.

\item
For model II, a lower bound on the mass $\mhp$ can be obtained from the
measured branching ratios of $B \to K^* \gamma$ decays:
\beq \mhp \gtrsim 200\ \ {\rm  or} \ \  300 {\rm GeV},
\eeq
if one uses $F_{K^*} =0.25 \pm 0.06$ or $F_{K^*} =0.38\pm 0.06$ in
the calculation, as illustrated by \fig{fig:fig2} and \fig{fig:fig4}. From
\fig{fig:fig3}, a lower limit of  $\tan{\beta} > 0.5$ can also be obtained
from the data.

\item
In the model III-B, the new physics contributions to $C_{7,8}(\mw)$
are larger than its SM counterparts in size and change the sign of the dominant
Wilson coefficient $C_7(m_b)$ from negative to positive, as given in
Eqs.(\ref{eq:a7ukmb-m3b}) and (\ref{eq:a7ckmb-m3b}).

\item
In the model III-B, the ranges of
\beq
226 \leq \mhp \leq 293 {\rm GeV}, \ \ {\rm and} \ \ \mhp \geq 1670 {\rm GeV},
\label{eq:lkac}
\eeq
are still allowed by the measured $\calb (B \to K^* \gamma)$ as given in
Eq.(\ref{eq:bvga-exp}).
From the experimental upper bounds on  $\calb (B \to \rho \gamma)$, we find the
lower limit on $\mhp$
\beq
\mhp \geq 206 {\rm GeV}
\label{eq:lra2}
\eeq
when  the $2\sigma$ theoretical errors are also taken into account. Above
limits on $\mhp$ are comparable with those obtained from the inclusive $B \to X_s \gamma$
decay \cite{xiao03}.

\item
The model III-B prediction for the isospin symmetry breaking of
$B \to K^* \gamma$ decay is $\Delta_{0-}(K^* \gamma) =
(-5.6 ^{+4.4}_{-2.8})\%$, which is small in size but has an opposite
sign with the measured value, as illustrated in Figs.~\ref{fig:fig7}
and \ref{fig:fig8}. A positive $C_7(m_b)$ is therefore disfavered by the measured
value of $\Delta_{0-}^{exp} (K^*\gamma) = (3.9 \pm 4.8)\%$, but still can not be excluded
if we take the large theoretical and experimental errors into account.

\item
The theoretical predictions for CP asymmetry $\acp(B \to K^* \gamma)$
is always less than one percent in magnitude in the SM and all three types of the general
2HDM's considered here.
For $B \to \rho \gamma$ decay, however, its CP asymmetry can be as large as about $10\%$ in
size in the SM and all three types of 2HDM's and have a strong dependence on
the variations of the scale $\mu={\cal O}(m_b)$ and the CKM angle $\gamma$, as
shown in Figs.~\ref{fig:fig12} and \ref{fig:fig13}. It is interesting
to see from \fig{fig:fig13} that the CP asymmetry in model III-B has an opposite sign
with the one in the SM. This feature may be used as a good observable
to distinguish the model III-B (or a positive $C_7(m_b)$) with the SM (a negative
$C_7(m_b)$).

\item
For $B \to \rho \gamma$ decay, the isospin symmetry breaking is less than $10\%$ in the
region of $\gamma = [ 40 \sim 70]^\circ$ as preferred by the global fit result \cite{pdg2003},
but can be as large as $20$ to $40\%$ in the regions of $\gamma \leq 10^\circ$ and
$\gamma \geq 120^\circ$, as can be seen clearly in Fig.~\ref{fig:fig14}.
The SM and model III-B predictions for isospin breaking have
an opposite sign for small or large values of the CKM angle $\gamma$.

\item
The U-spin symmetry breaking $\Delta U(K^*,\rho)$ in the SM and all 2HDM's considered here
is generally small in size: $\sim 10^{-7}$.

\end{itemize}


\begin{acknowledgments}

Z.J.Xiao is very grateful to the high energy section of ICTP, Italy, where part
of this work was done, for warm hospitality and financial support.
This work was supported by the National Natural Science Foundation of
China under Grant No.~10075013 and 10275035, and by the Research Foundation
of Nanjing Normal University under Grant No.~214080A916.

\end{acknowledgments}

\newpage

\begin{appendix}

\section{$G_i$ and $H_i^V$ functions}\label{app:gi}

In this Appendix, the explicit expressions or numerical values of  all
$G_i$ and $H_i^V$ functions appeared in Eq.(\ref{eq:avga}) will be listed.
For more details of these functions, one can see Ref.\cite{bosch02b} and
references therein.

\beq
G_1(z) &=& \displaystyle \frac{52}{81}\ln\frac{\mu}{m_b}
    +\frac{833}{972}-\frac{1}{4}[a(z)+b(z)]+\frac{10i\pi}{81}, \\
G_2(z) &=& \displaystyle -\frac{104}{27}\ln\frac{\mu}{m_b}
     -\frac{833}{162} +\frac{3}{2}[a(z) +b(z)]-\frac{20i\pi}{27}, \\
G_3      &=& \displaystyle \frac{44}{27}\ln\frac{\mu}{m_b}
    + \frac{598}{81} +\frac{2\pi}{\sqrt{3}} +\frac{8}{3} X_b
    -\frac{3}{4}a(1) +\frac{3}{2}b(1) +\frac{14i\pi}{27}, \\
G_4(z_c) &=& \displaystyle \frac{38}{81}\ln\frac{\mu}{m_b}
    + -\frac{761}{972} -\frac{\pi}{3\sqrt{3}} -\frac{4}{9} X_b
    +\frac{1}{8} a(1) +\frac{5}{4} b(z_c) -\frac{37i\pi}{81} \\
G_5      &=& \displaystyle \frac{1568}{27}\ln\frac{\mu}{m_b}
    + \frac{14170}{81} +\frac{8\pi}{\sqrt{3}} +\frac{32}{3} X_b
    -12 a(1) +24 b(1) +\frac{224i\pi}{27},       \\
G_6(z_c) &=& \displaystyle -\frac{1156}{81}\ln\frac{\mu}{m_b}
    + \frac{2855}{486} -\frac{4\pi}{3\sqrt{3}} -\frac{16}{9} X_b\non
    && -\frac{5}{2} a(1) +11 b(1) +9 a(z_c) +15 b(z_c) -\frac{574i\pi}{81}, \\
G_8      &=& \displaystyle \frac{8}{3}\ln\frac{\mu}{m_b}
    + \frac{11}{3} -\frac{2\pi^2}{9} +\frac{2i\pi}{3}, \\
\eeq
where
\beq
X_b &=& \int_0^1 \!dx \int_0^1 \!dy \int_0^1 \!dv x y \ln[v+x(1-x)(1-v)(1-v+v y)]
\approx -0.1684, \label{eq:xb}\\
  a(1) &\simeq & 4.0859 + \frac{4i\pi}{9},\label{eq:a1} \\
  b(1) &=& \frac{320}{81} - \frac{4 \pi}{3 \sqrt{3}} + \frac{632\pi^2}{1215}
  - \frac{8}{45} \left[ \frac{d^2 \ln \Gamma(x)}{dx^2} \right]_{x=\frac{1}{6}}+ \frac{4i\pi}{81}
  \simeq 0.0316 + \frac{4i\pi}{81}, \label{eq:b1}\\
a(z_u)&=& \left ( -1.93 + 4.96 i \right )\times 10^{-5}, \label{eq:azu}\\
a(z_c)&=&   1.525 + 1.242 i , \label{eq:azc}\\
b(z_u)&=& \left ( 1.11 + 0.28 i \right )\times 10^{-5}, \label{eq:bzu}\\
b(z_c)&=&  -0.0195 + 0.1318 i , \label{eq:bzc}
\eeq
where $z_q=m_q^2/m_b^2$ and the masses $m_q$ ($q=u,c,b$) as listed in \tab{input} have been
used to obtain the numerical results. The explicit analytical expressions for
$a(z)$ and $b(z)$ can be found for example in Ref.\cite{bosch02b}.

For the $H_i^V$ functions, we have
\beq
H^V_1(z_p)   &=& -\frac{2\pi^2}{9}\frac{f_B f^\perp_V \lambda_B}{F_V m_B}
  \int^1_0 dv\, h(\bar v,z_p) \Phi_V^\perp(v),   \label{eq:h1v}\\
H^V_2      &=& 0\\
H^V_3      &=& -\frac{1}{2}\left[ H^V_1(1) +H^V_1(0)\right], \\
H^V_4(z_c) &=&  H^V_1(z_c)-\frac{1}{2}H^V_1(1), \label{eq:h4v}\\
H^V_5      &=&  2 H^V_1(1), \label{eq:h5v}\\
H^V_6(z_c) &=& -H^V_1(z_c)+\frac{1}{2}H^V_1(1) =-H^V_4(z_c), \label{eq:h6v}\\
H^V_8   &=& -\frac{4\pi^2}{3}\frac{f_B f^\perp_V \lambda_B}{F_V m_B}
  \left ( 1 -\alpha_1^V + \alpha_2^V + \cdots \right ),   \label{eq:h8v}
\eeq
where the hard-scattering function $h(u,z)$ is given by
\beq
h(u,z)&=& \frac{4z}{u^2}\left\{  Li_2\!\left[ \frac{2}{1-\sqrt{\frac{u-4z+i\varepsilon}{u}}}\right ]
+ Li_2\!\left[ \frac{2}{1+\sqrt{\frac{u-4z+i\varepsilon}{u}}}\right ]\right \}
-\frac{2}{u},
\eeq
where $Li_2[x]$ is the dilogarithmic function, and the function $h(u,z)$ is real for $u \leq 4z$ and
develops an imaginary part for $u > 4z$.  The light-cone wave function $\Phi_V^\perp(v)$
takes the form of
\beq
\Phi_V^\perp (v) &=& 6 v (1-v) \left [ 1 +  \alpha_1^V(\mu) C_1^{3/2}(2v-1)
+ \alpha_2^V(\mu) C_2^{3/2}(2v-1) + \cdots \right ]
\eeq
where $C_1^{3/2}(x)=3x$, $C_2^{3/2}(x)=\frac{3}{2}(5x^2 -1)$.

\section{NLO coefficients at $\mu= \mw$ in general 2HDM's} \label{app-wc}

For the completeness, we list here the expressions of the NLO functions $W_{i,j}, M_{i,j}$
and $T_{i,j}$ ($i=7,8$ and $j=YY, XY$) at the matching scale
$\muw=\mw$ in the general two-Higgs-doublet models.
For more details see Ref.\cite{bg98}.

The NLO functions proportional to the term $|Y|^2$ are
\beq
 W_{7,YY}(y) & = & \frac{2y}{9}\left[ \frac{8y^3-37y^2+18y}{(y -1)^4}\,
  {\rm Li}_2 \left(\!1\! -\! \frac{1}{y} \right)
+\frac{3y^3+23y^2-14y}{(y -1)^5}\ln^2y \right. \non
&& \left.
+ \frac{21y^4-192y^3-174y^2+251y-50}{9(y-1)^5} \ln y \right.     \non
&   & \left.
 +\frac{-1202y^3+7569y^2-5436y+797}{108(y-1)^4}\right]
 \ - \frac{4}{9} \, E_H    ,
\eeq
\beq
W_{8,YY} (y)& = & \frac{y}{6} \left[ \frac{13y^3-17y^2+30y}{(y-1)^4}\,
 {\rm Li}_2  \left(\!1\! -\! \frac{1}{y} \right)
 -\frac{17y^2+31y}{(y-1)^5}\ln^2y \right. \non
 && \left.
 +\frac{42y^4+318y^3+1353y^2+817y-226}{36(y-1)^5}\ln y \right. \non
 && \left.
+\frac{-4451y^3+7650y^2-18153y+1130}{216(y-1)^4}\right]
 \ - \frac{1}{6} \, E_H  ,
\eeq
\beq
 M_{7,YY} (y)  & = &\frac{y}{27}\left[
  \frac{-14y^4+149y^3-153y^2-13y+31-(18y^3+138y^2-84y) \ln y}{(y-1)^5}
 \right], \\
 M_{8,YY} (y) & = &\frac{y}{36}\left[
 \frac{-7y^4+25y^3-279y^2+223y+38+(102y^2+186y) \ln y}{(y-1)^5}
            \right], \\
T_{7,YY} (y) &=& \frac{y}{9} \, \left[\frac{47y^3-63y^2+9y+7-(18y^3+30y^2-24y)
                       \ln y}{(y-1)^5} \right] ,        \\
T_{8,YY} (y)  &=& \frac{2y}{3} \left[\frac{-y^3-9y^2+9y+1+(6y^2+6y)
                       \ln y}{(y-1)^5} \right],
\label{eq:nlo-yy}
\eeq
with
\beq
E_H (y) & = & \frac{y}{36} \, \left[ \frac{7y^3-36y^2+45y-16+(18y-12) \ln y}{(y - 1)^4}
\right].
\eeq

The NLO functions proportional to the term $(XY^*)$ are
\beq
 W_{7,XY}(y) & = & \frac{4 y}{3}\left[  \frac{8y^2-28y+12}{3(y-1)^3} \,
 {\rm Li}_2 \left(\!1\! -\! \frac{1}{y} \right)+
\frac{3y^2+14y-8}{3(y-1)^4}\ln^2y \right. \non
& & \left. + \frac{4y^3-24y^2+2y+6}{3(y-1)^4}\ln y
+\frac{-2y^2+13y-7}{(y-1)^3}\right], \\
 W_{8,XY}(y) & = & \frac{y}{3} \left[  \frac{17y^2-25y+36}{2(y-1)^3}\,
 {\rm Li}_2 \left(\!1\! -\! \frac{1}{y} \right) -
  \frac{17y+19}{(y-1)^4}\ln^2y \right. \non
   &  & \left. +\frac{14y^3-12y^2+187y+3}{4(y-1)^4}\ln y
-\frac{3(29y^2-44y+143)}{8(y-1)^3}\right] ,  \\
 M_{7,XY}(y)& = & \frac{2 y}{9} \, \left[  \frac{-8y^3+55y^2-68y+21-(6y^2+28y-16) \ln y}{(y-1)^4}
 \right], \\
 M_{8,XY}(y) & = & \frac{y}{6} \, \left[ \frac{-7y^3+23y^2-97y+81 +(34y+38) \ln y}{(y-1)^4}
                    \right],   \\
T_{7,XY} (y)&=& \frac{2y}{3} \, \left[ \frac{13y^2-20y+7-(6y^2+4y-4)
                       \ln y}{(y-1)^4} \right],  \\
T_{8,XY}   &=& 2y \, \left[ \frac{-y^2-4y+5+(4y+2) \ln y}{(y-1)^4} \right] \, ,
\label{eq:nlo-xy}
\eeq
where $y=\mt^2/\mhp^2$.

\end{appendix}

\newpage

\newpage
\begin{thebibliography}{99}

\bibitem{buras96}
G.~Buchalla, A.J.~Buras, and M.E.~Lautenbacher, \rmp {\bf 68}, 1125 (1996).

\bibitem{hurth02}
For recent reviews of rare B decays, see T.Hurth, hep-ph/0212304;
M.~Battaglia  {\it et al.}, hep-ph/0304132.

\bibitem{bosch03}
S.W.~Bosch and G.~Buchalla, in {\em Proceedings of the Second Workshop on the CKM Unitarity
Triangle, IPPP Duram, April 2003}, edited by P.~Pall, J.~Flynn, P.~Kluit, and A.~Stocchi,
eConf C0304052:WG203 (2003).

\bibitem{jessop02}
C.~Jessop, {\em A world average for $B \to X_s \gamma$}, SLAC-PUB-9610.

\bibitem{cmm97}
K.G.~Chetyrkin, M.~Misiak, M.~Munz, \plb {\bf 400}, 206(1997),
{\bf 425}, 414(E) (1998).

\bibitem{kagan99}
A.L.~Kagan and M.~Neubert, \epjc {7}, 5 (1999).

\bibitem{buras02}
A.J.~Buras, A.~Czarnecki, M.~Misiak, and J.~Urban, \npb {\bf 611}, 488 (2001),
{\bf 631}, 219 (2002), and reference therein.

\bibitem{gm01}
P.~Gambino and and M.~Misiak, \npb {\bf 611}, 338 (2001).

\bibitem{greub03}
For recent devolopments see C.~Greub, talk  presented at the EPS-2003, 17 - 23 July 2003,
Aachen, Germany.

\bibitem{carena01}
M.~Carena, D.~Garcia, U.~Nierste and C.~E.~Wagner, \plb {\bf 499}, 141 (2001);
G.~Degrassi, P.~Gambino and G.~F.~Giudice, JHEP {\bf 0012}, 009 (2000);
G.~D'Ambrosio, G.~F.~Giudice, G.~Isidori and A.~Strumia, \npb {\bf 645}, 255 (2002).

\bibitem{bor00}
F.~Borzumati, C.~Greub, T.~Hurth and D.~Wyler, \prd {\bf 62}, 075005 (2000);
T.~Besmer, C.~Greub and T.~Hurth, \npb {\bf 609},  359 (2001).

\bibitem{2hdm}
S.~Glashow and S.~Weinberg, \prd {\bf 15}, 1958 (1977);
J.F.~Gunion, H.E.~Haber, G.~Kane, and S.~Dawson, {\em The Higgs Hunter's Guide},
Addison Wesley, Redwood-City (1990), and references therein.

\bibitem{xiao03}
Z.J.~Xiao and L.B.~Guo, hep-ph/0309103, \prd 68 (2003)in press, and references therein.

\bibitem{cleo-vg}
CLEO Collaboration, R.~Ammer {\it et al.}, \prl 71, 674 (1993);
T.~Coan {\it et al.}, \prl, 84, 5283 (2000).

\bibitem{babar-vg}
B.~Aubert {\it et al.}, BaBar Collaboration, \prl 88, 101805 (2002);
hep-ex/0306038;


\bibitem{belle-vg}
K.~Abe {\it et al.}, Belle Collab., {\em Measurement of the $B \to K^* \gamma$ branching
fraction and asymmetries}, Belle-Conf-0319, EPS-ID 537.

\bibitem{nakao03}
For recent developments, see M.~Nakao ( for Belle, BaBar, CLEO,VDF and D0 Collaboratios ),
 {\rm Radiative and electroweak rare B decays}, talk presented
 at LP 2003, Aug.12, 2003.

\bibitem{deshpande87}
N.G.~Deshpande, P.~Lo, and J.~Trampetic, \prl {\bf 59}, 183 (1987).

\bibitem{greub95}
C.~Greub, H.~Simma and D.~Wyler, \npb {\bf 434}, 39 (1995) [Erratum-ibid. B {\bf 444}, 447
(1995)].

\bibitem{aaw99}
H.H.~Asatryan, H.M.~Asatrian, and D.~Wyler, \plb {\bf 470}, 223 (1999).

\bibitem{li99}
H.-n.~Li and G.L. Lin, \prd {\bf 60}, 054001 (1999).

\bibitem{beneke01}
M.~Beneke, T.~Feldmann and D.~Seidel, \npb {\bf 612}, 25 (2001).

\bibitem{ali02a}
A.~Ali and A.Y.~Parkhomenko, \epjc {\bf 23}, 89 (2002).

\bibitem{bosch02a}
S.W.~Bosch and G.~Buchalla, \npb {\bf 621}, 459 (2002).

\bibitem{bbns99}
M.~Beneke, G.~Buchalla, M.~Neubert, and C.T.~Sachrajda, \prl {\bf 83}, 1914
(1999); \npb 591, 313(2000); \npb 606, 245 (2001).


\bibitem{bosch02b}
S.W.~Bosch,  { \em Exclusive Radiative Decays of B Mesons in QCD Factorization},
PH.D thesis,  hep-ph/0208203.

\bibitem{kagan02}
A.L.~Kagan and M.~Neubert, \plb  {\bf 539}, 227 (2002).


\bibitem{ali00}
A.~Ali, T.~Handoko, and D.~London, \prd {\bf 63}, 014014 (2001);
A.~Ali and E.~Lunghi, \epjc {\bf 26}, 195 (2002).

\bibitem{ckm}
M.~Kabayashi, T.~Maskawa, Prog. Theor. Phys. {\bf 49}, 652 (1973).

\bibitem{buras94}
A.J.~Buras, M.~Misiak, M.~M\"unz, and S.~Pokorski, \npb {\bf 424}, 374 (1994).

\bibitem{pdg2003}
Particle Data Group, K.~Hagiwara {\it et al.}, \prd {\bf 66}, 010001 (2002) and 2003 partial
update for edition 2004 (URL: http://pdg.lbl.gov).

\bibitem{ball98}
P.~Ball and V.M.~Braun, \prd {\bf 58}, 094016 (1998).

\bibitem{b03}
D.~Becirevic, talk given at the Ringberg Phenomenology Workshop on Heavy
Flavors, Ringberg Castle, Tegernsee, Germany, May 2003.

\bibitem{atwood97}
D.~Atwood, L.~Reina and A.~Soni, \prd {\bf 55}, 3156 (1997).

\bibitem{hou92}
T.P.~Cheng and M.~Sher, \prd {\bf 35}, 3484 (1987);
M.~Sher and Y.~Yuan, \prd {\bf 44}, 1461 (1991);
W.S.~Hou, \plb {\bf 296}, 179 (1992);
A.~Antaramian, L.J.~Hall,  and A.~Rasin, \prl {\bf 69}, 1871 (1992);
L.J.~Hall and S.~Winberg, \prd {\bf 48}, R979 (1993);
D.~Chang, W.S.~Hou,  and W.Y.~Keung, \prd {\bf 48}, 217 (1993);
Y.L.~Wu and L.~ Wolfenstein, \prl {\bf 73}, 1762 (1994);
D.~ Atwood, L.~Reina and A.~Soni, \prl {\bf 75}, 3800 (1995).


\bibitem{lo2hdm}
W.S.~Hou and R.S.~Willey, \plb {\bf 202}, 59 (1988);
S.~Bertolini {\it et al.}, \npb {\bf 353}, 591 (1991);
C.D.~ L\"u, \npb {\bf 441}, 33 (1994).

\bibitem{bg98}
F.M.~Borzumati and C.~Greub, \prd {\bf 58}, 074004 (1998);
{\bf 59}, 057501 (1999) (Addendum).

\bibitem{ciuchini98}
M.~Ciuchini, G.~Degrassi, P.~Gambino, and G.F.~Giudice, \npb {\bf 527}, 21
(1998).

\bibitem{crs98}
P.~Ciafaloni, A.~Romanino, and A.~Strumia, \npb {\bf 524}, 361 (1998).

\bibitem{aliev99}
T.M.~Aliev and E.O.~Iltan, \jpg {\bf 25}, 989 (1999).

\bibitem{chao99}
D.B.~ Chao, K.~heung, and W.Y.~ Keung, \prd {\bf 59}, 115006 (1999).

\bibitem{wu99}
K.~Kirs, A.~Soni, and G.H.~Wu, \prd {\bf 59}, 096001 (1999);
{\it ibid}, {\bf 62}, 116004 (2000); G.H.~Wu and
A.~Soni, {\it ibid}, {\bf 62}, 056005 (2000);


\bibitem{xiao01}
Z,J.~Xiao, C.S.~Li, and K.T.~Chao, \prd {\bf 63}, 074005 (2001);
J.J.~Cao, Z.J.~Xiao,  and G.R.~Lu, \prd {\bf 64}, 014012 (2001);
D.~Zhang, Z.J.~Xiao,  and C.S.~Li, \prd {\bf 64}, 014014 (2001);
Z.J.~Xiao, K.T.~Chao,  and C.S.~Li, \prd {\bf 65}, 114021 (2002).

\bibitem{gambino01}
P.~Gambino, \jpg {\bf 27}, 1199 (2001).

\bibitem{isosm}
M.~Gronau and J.L.~Rosner, \plb {\bf 500}, 247 (2001);
M.~Gronau, \plb {\bf 492}, 297 (2000);
T.~Hurth and T.~Mannel, \plb {\bf 511}, 196 (2001);
R.~Fleischer, \plb {\bf 459}, 306 (1999).

\end{thebibliography}
\end{document}